\documentclass[twocolumn,showpacs,aps]{revtex4}
\usepackage{bm}

\begin{document}
\title{Relaxed Bell inequalities and Kochen-Specker theorems}
\author{Michael J. W. Hall}
\affiliation{Theoretical Physics, Research School of Physics and Engineering,  Australian National
University, Canberra ACT 0200, Australia}

\begin{abstract}
The combination of various physically plausible properties, such as no signaling, determinism, and experimental free will, is known to be incompatible with quantum correlations.  Hence, these properties must be individually or jointly relaxed in any model of such correlations.  The necessary degrees of relaxation are quantified here via natural distance and information-theoretic measures. This allows quantitative comparisons between different models in terms of the resources, such as the number of bits of randomness, communication, and/or correlation, that they require.   For example, measurement dependence is a relatively strong resource for modeling singlet state correlations, with only 1/15 of one bit of correlation required between measurement settings and the underlying variable. It is shown how various `relaxed' Bell inequalities may be obtained, which precisely specify the complementary degrees of relaxation required to model any given violation of a standard Bell inequality. The robustness of a class of Kochen-Specker theorems, to relaxation of measurement independence, is also investigated.  It is shown that a theorem of Mermin remains valid unless measurement independence is relaxed by 1/3.  The Conway-Kochen `free will' theorem and a result of Hardy are less robust, failing if measurement independence is relaxed by only 6.5\% and 4.5\%, respectively. An appendix shows the existence of an outcome independent model is equivalent to the existence of a  deterministic model.
\end{abstract}

\maketitle
\section{Introduction}

Bell inequalities and Kochen-Specker theorems demonstrate that at least one very plausible property - such as no signaling, determinism, or measurement independence - does not hold in a world that exhibits quantum correlations \cite{bell, chsh,bell1,jointprob,bellfact,collins,ks,bellrev,heywood,mermin,hardy,conway}.  Any model or simulation of quantum systems must, therefore, give up at least one such property. But {\it how much} must be given up ?  Is $20\%$ indeterminism sufficient to maximally violate a Bell inequality?  Is a combination of 5\% signaling and 10\% measurement dependence enough to simulate singlet state correlations?  

The question of the degree to which such properties must be relaxed is of fundamental interest in constructing physical theories. It is also relevant to understanding so-called `quantum nonlocality' as a physical resource, in tasks such as quantum computation and secure quantum cryptography.  For example, singlet state correlations can be modeled by giving up  100\% of determinism  \cite{branc},  or 14\% of measurement independence (related to the freedom to choose experimental settings) \cite{free}.  Hence, indeterminism appears to be a weaker `nonlocal' resource than experimental free will, for simulating the  singlet state.

The main aim of this paper is to carefully define and quantify the degrees to which certain physical properties hold for a given model of correlations, and show how these may be applied to determine (i) optimal singlet state models; (ii) the minimal degrees of relaxation required to simulate violations of various Bell inequalities, and (iii) the relative robustness of Kochen-Specker theorems.  

The physical properties considered are precisely those which are brought into question by the existence of quantum correlations. The quantitative nature of the results helps considerably to clarify the nature of these correlations, as well the resources required for their simulation.

The general form of underlying (or `hidden variable') models of statistical correlations is recalled in Sec.~II, and the degrees to which such underlying models possess a number of physically plausible properties, such as determinism, outcome independence, no signaling and measurement independence, are defined and discussed in Secs.~III-V.  Both statistical and information-theoretic based measures are considered.  These sections, together with Appendix A, also demonstrate that the properties of determinism and outcome independence are effectively equivalent, and relate the degree of communication required to implement a given nonlocal model to the amount of signaling permitted by the model.   

In Sec.~VI it is demonstrated that there are three canonical models of singlet state correlations, corresponding to the minimal degrees to which one of the above mentioned properties must be relaxed while maintaining the others.  The corresponding information-theoretic resources required are 1 bit of randomness generation or outcome correlation, 1 bit of signaling or communication, and $1/15$ of one bit of correlation between the underlying variable and the measurement settings.

It is shown in Sec.~VII, together with Appendices B and C, how to derive `relaxed' Bell inequalities.  These precisely quantify the individual and/or joint degrees of relaxation required to model a given violation of a standard Bell inequality. Examples include the joint relaxation of determinism, no signaling and measurement independence for the Bell-CHSH inequality \cite{chsh}, verifying a recent conjecture \cite{bis}; the relaxation of outcome independence for the same inequality; and the relaxation of indeterminism and no signaling for a form of the $I_{3322}$ inequality \cite{collins}.  

Sec.~VIII shows how local deterministic models may be obtained for the perfect correlations underlying members of a strong class of  Kochen-Specker theorems \cite{heywood,mermin,hardy,conway}.  These models require the relaxation of measurement independence, and the minimal degree of relaxation quantifies the relative robustness of such theorems.  It is found that a version due to Mermin \cite{mermin} is the most robust, requiring relaxation by 1/3.

Conclusions are given in Sec.~IX.

\section{Underlying models}

Consider a given set of statistical correlations, $\{p(a,b|x,y)\}$, where the pair $(a,b)$ labels the possible outcomes of a joint experiment $(x,y)$,  for some fixed preparation procedure.   Any underlying model of these correlations introduces an underlying variable $\lambda$ on which the correlations depend, which is typically interpreted as representing information about the preparation procedure.  From Bayes theorem one has the identity
\begin{equation} \label{underlying}
p(a,b|x,y) = \int d\lambda\, p(a,b|x,y,\lambda)\, p(\lambda|x,y) ,
\end{equation}
with integration replaced by summation over any discrete ranges of $\lambda$.  
A given underlying model specifies the type of information encoded by $\lambda$, and the underlying probability densities $p(a,b|x,y,\lambda)$ and $p(\lambda|x,y)$.

For example, the standard Hilbert space model of quantum correlations represents the underlying variable by a density operator, $\rho$, and the joint measurement setting by a probability operator measure, $\{ E^{xy}_{ab} \}$, with
\begin{equation} \label{qcorr}
 p(a,b|x,y,\rho) = {\rm tr}[\rho E^{xy}_{ab}],~~~p(\rho|x,y) = \delta(\rho-\rho_0).  
 \end{equation}
One may alternatively use a pure state model, of the form
\[ p(a,b|x,y,\psi) = \langle\psi| E^{xy}_{ab}|\psi\rangle,~~~p(\psi|x,y) = p_0(\psi), \]
where $\lambda$ is restricted to the set of unit vectors $\{\psi\}$ on the Hilbert space, and the models are related by $\rho_0\equiv \int d\psi\, p_0(\psi)|\psi\rangle\langle\psi|$.

A given underlying model may or may not satisfy various physically plausible properties, such as no signaling, determinism, outcome independence, etc.  The violation of Bell inequalities and Kochen-Specker theorems, by certain quantum correlations, implies that at least one such property must be relaxed by any model of these correlations.  The necessary degrees of relaxation are the central concern of this paper, and help both to clarify and quantify the nonclassical nature of quantum entanglement.  

These properties are defined in Secs.~III-V below, and natural measures of the degree to which they hold, for a given model, are defined.  These measures can generally be expressed in terms of the variational distance between two probability distributions $P$ and $Q$, 
\[
D(P,Q) := \sum_n |P(n) - Q(n)| ,
\]
or in terms of Shannon entropy and mutual information.
While the distance measures are typically easier to work with, the information-theoretic measures have the advantage of directly quantifying various resources, such as randomness, correlation information, and communication capacity.

\section{Determinism and outcome independence}

\subsection{Physical significance}

Determinism is the property that all outcomes can be predicted with certainty, given knowledge of the underlying variable $\lambda$, i.e., 
$p(a,b|x,y,\lambda) = 0 ~~{\rm or}~~1.$
This is easily shown to be equivalent to the property that all underlying {\it marginal} probabilities are deterministic, i.e., to
\begin{equation} \label{det}
p(a|x,y,\lambda),\, p(b|x,y,\lambda) \in \{0,1\} . 
\end{equation}
In contrast, outcome independence is the property that, given knowledge of the underlying variable $\lambda$, the joint measurement outcomes are uncorrelated \cite{jarrett}, i.e., 
\begin{equation} \label{out}
p(a,b|x,y,\lambda) = p(a|x,y,\lambda)\, p(b|x,y,\lambda) . 
\end{equation}
Thus, any observable correlations arise only as a consequence of ignorance of the underlying variable.  

Any deterministic model is trivially outcome independent (see Appendix A), and so it may appear that determinism is a more restrictive property.  However, as shown in Appendix A, the difference between these two properties is largely cosmetic: {\it for any set of statistical correlations, $\{p(a,b|x,y)\}$, there exists an underlying deterministic model ${\cal M}$ if and only if there exists an underlying outcome independent model ${\cal M'}$}.  Furthermore, ${\cal M}$ satisfies no-signaling or measurement independence if and only if ${\cal M'}$ does.  

At least two plausible arguments may be made for the existence of an underlying deterministic (and hence outcome independent) model of physical correlations.  
The first is based on a `realist' interpretation of probability, in which the assignation of probabilities to measurement outcomes merely reflects ignorance as to an underlying `real state of affairs'.  This implies an underlying deterministic model for the outcomes, where $p(\lambda|x,y)$ in Eq.~(\ref{underlying}) describes ignorance of the precise state of affairs.   

This argument is easily countered by adopting a non-realist interpretation of probability, with measurement considered to be an act of creation rather than one of revelation \cite{prob,bohr}. Indeed, Bohr stated that ``we have in each experimental arrangement \dots not merely to do with the ignorance of the value of certain physical quantities, but with the impossibility of defining these quantities in an unambiguous way'' \cite{bohr}.  For example, one may  adopt a Bayesian interpretation of probability, where probabilities reflect consistent methods for making predictions on the basis of given knowledge \cite{bayes}, without requiring the existence of some underlying `perfect' knowledge.

The second main argument for determinism is based on the existence of perfect correlations.  In particular, as first pointed out by Einstein, Podolsky and Rosen \cite{epr}, perfect quantum correlations can exist between the outcomes corresponding to a given joint measurement setting $(x,y)$.  Thus, knowledge of the outcome for setting $x$ immediately implies knowledge of the outcome for setting $y$, and vice versa.  If no signaling between the two measurement regions is permitted, it immediately appears that the outcomes must have been predetermined - how else could such a perfect correlation be realised ?  Since quantum mechanics does not assign deterministic values to these outcomes, some underlying model must then do so.  This argument was also used by Bell in obtaining the original Bell inequality \cite{bell}.  

However, this argument may also be countered, even when no signaling is assumed.  For example, in the many-worlds interpretation of quantum mechanics, the two observers may in fact obtain random outcomes that do {\it not} always satisfy the predicted correlation - in which case they will simply end up in different branches of the universal wave function, unable to compare their inconsistent results \cite{manyworlds}.  In Bayesian interpretations, the rebuttal is that the correlations are a property of degrees of belief of observers (which may be informed by quantum models), rather than of some physical state {\it per se}, where any knowledge gained about one outcome from the other outcome (eg, due to a perfect correlation) merely reflects a local and consistent updating of either observer's degree of belief \cite{bayes}.

\subsection{Indeterminism and outcome dependence}

The degree of {\it indeterminism} of an underlying model may be defined as just how far away the marginal probabilities can be from the deterministic values of 0 and 1 in Eq.~(\ref{det}).  This is the smallest positive number, $I$, such that 
\begin{equation} \label{idef}
p(a|x,y,\lambda),\, p(b|x,y,\lambda)\in [0,I]\cup [1-I,1] .
\end{equation} 
Thus, $0\leq I\leq 1/2$, with $I=0$ if and only if the probabilities are confined to $\{0,1\}$ as per Eq.~(\ref{det}), i.e., if and only if the model is deterministic \cite{bis,inote}.  

A simple measure of outcome dependence, $O$, is the maximum variational distance between an underlying joint distribution and the product of its marginals, i.e., 
\begin{equation} \label{odef}
O:= \sup_{x,y,\lambda} \sum_{a,b} {\big|} p(a,b|x,y,\lambda) - p(a|x,y,\lambda)\, p(b|x,y,\lambda){\big|} .
\end{equation}
Thus, $0\leq O\leq 2$, and it follows immediately from Eq.~(\ref{out}) that $O=0$ if and only if outcome independence is satisfied.  

As noted above, the properties of determinism and outcome independence are closely related.  For example, as shown in Appendix A, for the particular case of two-valued outcomes one has the tight inequality
\begin{equation} \label{oi}
O \leq 4I(1-I) \leq 1 .
\end{equation}
This inequality chain is saturated, for example, by the singlet state of two qubits (see Sec.~VI), and by PR-boxes \cite{pr}. In both cases one has the maximum possible degrees of indeterminism  and outcome dependence, i.e., $I=1/2$ and $O=1$.

\subsection{Random bits and outcome correlation}

Indeterminism corresponds to a degree of randomness.  Hence, a natural information-theoretic measure of indeterminism is given by the maximum entropy of the underlying marginal probability distributions:
\begin{equation} \label{crand}
C_{random} := \sup_{x,y,\lambda} \{ H_{x,y,\lambda}(A), H_{x,y,\lambda}(B) \} ,
\end{equation}
where $H_{x,y,\lambda}(A)$ denotes the Shannon entropy of the outcome distribution $\{p(a|x,y,\lambda)\}$.  Thus, $C_{random}$ is the maximum number of random bits that must be generated to simulate a local outcome distribution, and $C_{random}=0$ for deterministic models.  Since there is an underlying marginal probability arbitrarily close to $I$, one has the lower bound
\begin{equation} \label{crandbound}
C_{random} \geq  h(I) ,
\end{equation}
with equality for the case of two-valued outcomes, where
\begin{equation} \label{hx}
h(x):=-x\log_2 x-(1-x)\log_2(1-x) .
\end{equation}

A corresponding information-theoretic measure of outcome dependence is given by the maximum Shannon mutual information between the outcomes:
\begin{eqnarray} \label{cresidual}
C_{outcome} &:=& \sup_{x,y,\lambda} H_{x,y,\lambda}(A:B) \\ \nonumber
= \sup_{x,y,\lambda} &\sum_{a,b}& p(a,b|x,y,\lambda) \log_2 \frac{p(a,b|x,y,\lambda)}{p(a|x,y,\lambda)p(b|x,y,\lambda)} .
\end{eqnarray}
This quantifies the maximum degree of correlation that is present between measurement outcomes, given knowledge of the underlying variable $\lambda$ \cite{infnote}, and vanishes for models satisfying outcome independence via Eq.~({\ref{out}).  

One has the relations
\begin{equation} \label{pinsker}
 C_{random}\geq C_{outcome} \geq \frac{1}{2} O^2 \log_2 e ,
 \end{equation}
where the upper bound follows from Eq.~(\ref{crand}), and the (nontight) lower bound from Pinsker's inequality \cite{pinsker}.  For the case of two-valued measurement outcomes this lower bound can be improved to the tight bound
\begin{equation} \label{obound}
C_{outcome} \geq 1 - h(\frac{1+O}{2}),
\end{equation}
in analogy to Eq.~(\ref{crandbound}).
In the standard Hilbert space model of singlet state spin correlations the maximum possible values for two-valued outcomes, $C_{random}=C_{outcome}= 1$ bit, are achieved (see Sec.~VI).  

\section{No signaling}

\subsection{Physical significance}

The property of no signaling (or parameter independence) is satisfied if the underlying marginal distribution associated with one setting is independent of the other setting, i.e., if
\begin{equation} \label{nosig}
p(a|x,y,\lambda) = p(a|x,y',\lambda),~p(b|x,y,\lambda) = p(a|x',y,\lambda)
\end{equation}
for all joint settings $(x,y)$, $(x,y')$ and $(x',y)$ of the model.  Thus, neither observer can affect the underlying measurement statistics of the other, via their choice of measurement setting.  Hilbert space models satisfy this property when the measure in Eq.~(\ref{qcorr}) has the tensor product form $E^{xy}_{ab}=E^x_a\otimes E^y_b$.

There are two strong arguments for requiring physical models to have the no signaling property. The first applies when the respective measurement settings are made in spacelike separated regions: altering the underlying statistics of a measurement in one such region, via varying a measurement setting in the other region, would violate the principle of relativistic causality and thus lead to the need to resolve various paradoxes.  

The second argument is that any signaling model underlying quantum correlations would have to explain the apparent `conspiracy' that quantum correlations are themselves nonsignaling.  In particular, all nonzero shifts in the underlying probability distributions, for any such underlying model, would have to average out to zero at the observable level.

However, while relativistic causality is a natural assumption, it still may be possible to consistently resolve apparent paradoxes if it does not hold.  Furthermore, it is often possible to transform  `conspiracies' into  well motivated `physical principles'. Thus, for example, in the deBroglie-Bohm model of quantum mechanics one can either postulate a typical universal initial state \cite{vigier}, or the existence of suitably smooth intial conditions relative to some degree of coarse graining \cite{westman}.   

\subsection{Signaling}

The degree of signaling is quite simply defined as the maximum possible shift in an underlying marginal probability for one observer, as the consequence of changing the measurement setting of the other observer.  More formally, one-way degrees of signaling are defined by \cite{bis}
\begin{eqnarray*} 
S_{1\rightarrow 2} &:=& \sup_{\{x,x',y,b,\lambda\} } \left| p(b|x,y,\lambda) - p(b|x',y\lambda) \right|,\\ 
S_{2\rightarrow 1} &:=& \sup_{\{x,y,y',a,\lambda\} } \left| p(a|x,y,\lambda) - p(a|x,y'\lambda) \right| ,
\end{eqnarray*}
where  $a$ and $b$ label measurement outcomes corresponding to measurement settings $x$ and $y$, respectively. Thus, for example, $S_{1\rightarrow 2}$ is the maximum possible shift in an underlying marginal probability distribution for the second observer, induced via changing a measurement setting of the first observer. If $S_{1\rightarrow 2}>0$ and $\lambda$ is known, the first observer can in principle communicate to the second observer merely by modulating the local measurement setting. 

The overall degree of signaling, for a given underlying model, is defined by
\begin{equation} \label{sig}
S:= \max \{S_{1\rightarrow 2},S_{2\rightarrow 1}\}  .
\end{equation}
It follows that $0\leq S\leq 1$, and $S=0$ for nonsignaling models \cite{varnote}.  

The degrees of indeterminism and signaling, $I$ and $S$, are not fully independent of one another.  For example, in a deterministic model the underlying marginal probabilities are restricted to the values $0$ and $1$, and hence only a probability shift of unity is possible between these values.  More generally, any shift $S$ in a marginal probability value must keep it in the range $[0,I]\cup[1-I,1]$, i.e., the value must either stay in the same subinterval ($S\leq I$), or cross the gap between the  subintervals ($S\geq 1-2I$). Hence, 
\begin{equation} \label{imin}
I\geq \min \{ S, (1-S)/2 \} .
\end{equation}
In contrast, the degree of outcome dependence, $O$, is completely independent of $S$.

\subsection{Signaling capacity}

The maximum signaling capacity of a given model is, in analogy to Eq.~(\ref{sig}), given by
\begin{equation} \label{csig}
C_{sig} := \sup_{\lambda,x,y} \{ H_{x,\lambda}(A:Y), H_{y,\lambda}(B:X) \},
\end{equation}
where $H_{x,\lambda}(A:Y)$ denotes the Shannon mutual information between the  measurement outcome of the first observer and the measurement setting of the second observer, for fixed $x$ and $\lambda$.  Thus, $C_{sig}$  directly quantifies the amount of information that may be transmitted between observers via appropriate choices of measurement settings \cite{infnote}.   

The two measures $S$ and $C_{sig}$ are related via \cite{bis}
\begin{equation} \label{siginf}
C_{sig} \geq 1 - h(\frac{1+S}{2}),
\end{equation}
analogous to Eqs.~(\ref{crandbound}) and (\ref{obound}).
Thus, nonlocal communication is always possible, in principle, if $S>0$.

For example, the standard Hilbert space model in Eq.~(\ref{qcorr}) is nonsignaling, with $S=C_{sig}=0$.  On the other hand, for the deterministic Toner-Bacon model of the singlet state \cite{toner}, one has $S=1$, since the probability of one observer's outcome can flip between 0 and 1, in dependence on the choice of measurement made by the first observer.  Noting that the right hand side of Eq.~(\ref{csig}) cannot be greater than than 1 for two-valued measurements, it follows via Eq.~(\ref{siginf}) that $C_{sig}=1$ bit for this model.

\subsection{Relation to communication models}
 
The signaling capacity of a model is, {\it prima facie}, a different concept to the degree of nonlocal communication required to simulate a given model.  The signaling capacity is the amount of information which the observers are able to exploit, in principle, for arbitrary communication once the model is in place.  In contrast, the communication capacity may be defined as the amount of information required to be transmitted between observers to simulate the model.  The connections between the two concepts are explored and clarified below, in the context of one-way communication models.

In a one-way communication model, a message $m$ is communicated from the first observer to the second observer, which may depend on the measurement setting $x$ and a shared underlying variable $\lambda$ \cite{cov}.  The message is used to generate outcomes for the second observer, such that Eq.~(\ref{underlying}) is satisfied.  

For example, in the Toner-Bacon model of the singlet state one has   \cite{toner}
\[ m=f(x,\lambda) := ({\rm sgn}\,x\cdot\lambda_1)\,({\rm sgn}\,x\cdot\lambda_2) ,\]
where the underlying variable $\lambda\equiv (\lambda_1,\lambda_2)$ comprises two unit vectors $\lambda_1$ and $\lambda_2$ uniformly distributed over the unit sphere. The corresponding measurement outcomes are deterministically generated as $a=-{\rm sgn}\,x\cdot\lambda_1 $ and $b={\rm sgn}\,y\cdot(\lambda_1+m\lambda_2)$, for spin directions $x$ and $y$. 

Since $\lambda$ is known by both observers, the maximum information obtainable from $m$, about the measurement setting and outcome of the first observer, is given by the mutual information $H_\lambda(M:X,A)$. Since $m$ is the only communication used to generate the underlying correlations,  this information must subsume any information obtainable from the outcome $b$ for any measurement setting $y$ of the second observer.  Hence, 
\begin{equation} \label{red}
 H_\lambda(M:X,A) \geq \sup_y H_{y,\lambda}(B:X,A) \geq \sup_y H_{y,\lambda}(B:X).  
\end{equation}
The communication model will be said to be nonredundant if strict equality holds.

The communication capacity is defined to be the maximum possible mutual information that is communicated about $x$ and $a$ via the message $m$, i.e., 
\begin{equation} \label{ccom}
C_{commun} := \sup_\lambda H_\lambda(M:X,A) .
\end{equation}
It follows immediately via Eqs.~(\ref{csig}) and (\ref{red}), recalling the communication is one-way only, that
\begin{equation} \label{comsig}
C_{commun} \geq C_{sig},
\end{equation}
with equality for nonredundant models.

For a deterministic communication model (such as the Toner-Bacon model), the message and the outcome of the first observer are completely specified by $x$ and $\lambda$, i.e., 
\[ p(m,x,a|\lambda)  = \delta_{m,f(x,\lambda)}\delta_{a,\alpha(x,\lambda)}\,p(x|\lambda), \]
for suitable functions $f$ and $\alpha$. Hence, $H_\lambda(M:X,A)=H_\lambda(M)$, and Eq.~(\ref{ccom}) simplifies to 
\begin{equation} \label{detcom}
C_{determ~commun} = \sup_\lambda H_\lambda(M) 
\end{equation}
for such models, i.e., the communication capacity is just the maximum possible entropy of the message.

As an example, consider the Toner-Bacon model described above.  If the distribution of measurement settings of the first observer, $p(x)$, is uniformly distributed, then  $H_\lambda(M)=h(\pi^{-1}\cos^{-1}\lambda_1\cdot\lambda_2)$, with $h(x)$ defined as in Eq.~(\ref{hx}) \cite{toner}. This is equal to 1 bit for $\lambda_1\cdot\lambda_2=0$.  This is the maximum possible entropy $H_\lambda$ in Eq.~(\ref{detcom}), since $m$ only takes two values. Hence, 
\[ C^{TB}_{commun} = 1 ~{\rm bit} \]
for this model.  Note this also follows from Eq.~(\ref{comsig}), since $C_{sig}=1$ from the previous section.  An example of an indeterministic communication model is discussed in Sec.~VII~A.

Toner and Bacon have numerically calculated the average of $H_\lambda(M)$ over $\lambda$, for the case of a uniform distribution $p(x)$, as $\approx 0.85$ bits.  As a consequence of the deterministic nature of the model, one further finds
\begin{equation} \label{hml}
 H(M,\Lambda:X) = \langle H_\lambda(M)\rangle  \approx 0.85 {\rm ~bits} 
 \end{equation}
for this case. In contrast, $H(M:X)=0$ whenever the first observer's setting is independent of $\lambda$, i.e., $p(x|\lambda)=p(x)$, implying no information can be gained about this setting from the knowledge of $m$ alone.  

\section{Measurement independence and experimental free will}

\subsection{Physical significance}

Measurement independence is the property that the distribution of the underlying variable is independent of the measurement settings, i.e., 
\begin{equation} \label{free}
p(\lambda|x,y) = p(\lambda|x',y') 
\end{equation}
for any joint settings $(x,y)$, $(x',y')$.  It is trivially satisfied by the quantum model in Eq.~(\ref{qcorr}).  It follows immediately via Bayes theorem that this property is equivalent to each of 
\[ p(x,y|\lambda) = p(x,y),~~~~p(x,y,\lambda)=p(x,y)\,p(\lambda) ,\]
whenever there is a well defined distribution, $p(x,y)$, of joint measurement settings \cite{pxyexist}.  

Measurement independence, particularly in the form $p(x,y|\lambda) = p(x,y)$, is often justified by the notion of `experimental free will', i.e., that experimenters can freely choose between different measurement settings irrespective of the underlying variable $\lambda$ describing the system.  More neutrally, if random number generators are used to determine the measurement settings, it may be argued that the physical operation of these generators should be independent of the underlying variables describing the system that is to be measured.  

However, there is no {\it a priori} physical reason why the behaviour of experimenters or random generators should {\it not} be statistically correlated with a given system to some degree, reflecting a common causal dependence on some underlying variable.  For example, as has been clearly pointed out in the quantum context by Brans \cite{brans}, any fundamental deterministic model underlying nature should certainly predict the joint measurement settings (which are, after all, physical phenomena), to the same degree as it predicts the measurement outcomes.  

Further, a violation of measurement independence is not automatically inconsistent with apparent experimental freedom.  For example, suppose two experimenters run a series of experiments where they aim to choose their joint measurement settings according to some predetermined joint probability distribution $p(x,y)$.  For example, they might use random number generators to choose between local settings according to some factorisable joint distribution $p(x,y)=p(x)\,p(y)$.  It might be argued that an underlying correlation, between the joint settings and some underlying variable $\lambda$, could prevent such a pre-arranged joint distribution from being realised.  However, this is not so: such a realisation merely restricts the joint distribution of $x$, $y$ and $\lambda$ to be
\begin{equation} \label{pxyl}
 p(x,y,\lambda) = p(\lambda|x,y)\, p(x,y), 
 \end{equation}
irrespective of whether or not measurement independence is satisfied. 

Finally, it may be mentioned that the violation of measurement independence is natural for retrocausal models, in which future measurement settings may influence the past statistics of the underlying variable.   While retrocasuality is counter-intuitive in allowing two directions of time, Price has shown it is surprisingly robust to paradoxes \cite{price}.  However, of course, one does not require retrocausality to violate the measurement independence property in Eq.~(\ref{free}) \cite{brans}.

\subsection{Measurement dependence and correlation}

The degree to which an underlying model violates measurement independence is most simply quantified by the variational distance \cite{free}
\begin{equation} \label{mdep}
M:= \sup_{x,x',y,y'} \int d\lambda \,\left|p(\lambda|x,y) - p(\lambda|x',y')\right| .
\end{equation}
Thus, $M=0$ when Eq.~(\ref{free}) holds.  In contrast,  a maximum value of $M=2$ implies that there are at least two particular joint measurement settings, $(x,y)$ and $(x',y')$, such that for any physical state $\lambda$ at most {\it one} of these joint settings is possible.  Hence, the observers can exercise no experimental free will whatsoever to choose between the joint settings in this case.  Such a model has been given by Brans for any state of two qubits, where the underlying variable $\lambda$ in fact completely determines the joint measurement settings \cite{brans} (this model easily generalises to any set of statistical correlations).  Individual degrees of measurement dependence, $M_1$ and $M_2$, may also be defined for each observer \cite{free}, but will not be considered here.

The fraction of measurement independence corresponding to a given model is defined by \cite{free}
\begin{equation} \label{f}
F:= 1-M/2 .
\end{equation}
Thus, $0\leq F\leq 1$, with $F=0$ corresponding the case where no experimental free will can be exercised to choose between two particular settings.  Note that, geometrically, $F$ also represents the minimum degree of overlap between any two underlying distributions $p(\lambda|x,y)$ and $p(\lambda|x',y')$.

A natural information-theoretic characterisation of the degree of measurement dependence has been recently proposed by Barrett and Gisin \cite{barrett}.  In particular, the mutual information between the measurement settings and the underlying variable,
\[   H(X,Y:\Lambda) = \sum_{x,y}\int d\lambda\, p(x,y,\lambda)\log_2 \frac{p(x,y,\lambda)}{p(x,y)\, p(\lambda)} ,   \]
quantifies the degree of correlation between the joint measurement setting and the underlying variable \cite{infnote}.  It is well-defined whenever the joint distribution $p(x,y)$ exists \cite{pxyexist}, with $p(x,y,\lambda)$ given by Eq.~(\ref{pxyl}). 

For models satisfying measurement independence, there is no correlation and the mutual information vanishes via Eq.~(\ref{free}).  In contrast, for the Brans model of two qubits \cite{brans}, where the hidden variable uniquely determines the joint measurement setting, there is perfect correlation, and the mutual information can become infinitely large (eg, for the case of randomly chosen settings with $p(x,y)=1/(4\pi)^2$).

The {\it measurement dependence capacity} of a given model may be defined by maximising the mutual information over all possible distributions of measurement settings:
\begin{equation} \label{cdep}
  C_{meas~dep} := \sup_{p(x,y)} H(X,Y:\Lambda) .
\end{equation}  
Barrett and Gisin have shown the existence of deterministic nonsignaling models of the singlet state with $C_{meas~dep}\leq 1$ bit \cite{barrett}.  It will be shown in the following section that a recently proposed model of this type has $C_{meas~dep}=0.0663$ bits, i.e., no more than $\approx 1/15$ of one bit of mutual information is required to reproduce all spin correlations, for any distribution $p(x,y)$ of experimental settings.

\section{Minimal singlet state models}

To indicate how the above introduced measures allow quantitative comparisons between different models, three fundamental models of the singlet state correlations
\begin{equation} \label{singlet}
 p(a,b|x,y)  = \frac{1}{4}\left( 1 - ab \,x\cdot y\right) 
\end{equation}
are briefly examined here, where $a,b=\pm 1$ denote spin-up and spin-down outcomes for measurements in directions $x$ and $y$ respectively.    

Each of the three models corresponds to the minimum possible relaxation of one of the properties of determinism, outcome independence, no signaling, and measurement independence, while retaining the others.  It will be seen that measurement dependence is a particularly strong resource for modeling quantum correlations.

\subsection{Relaxing determinism}

First, consider the class of singlet state models which only relax determinism and/or outcome independence, i.e., for which $S=M=0$. The canonical member of this class is the standard Hilbert space model.  As noted in Sec.~III, this model has the maximum possible degrees of indeterminism and outcome dependence, 
\begin{equation} \label{ihs}
I^{HS}=1/2,~~~~O^{HS}=1, 
\end{equation}
as well as the maximum possible number of locally generated random bits and outcome correlation,  
\begin{equation} \label{chs}
C^{HS}_{random} = C^{HS}_{outcome} =1 {\rm ~bit}   . 
\end{equation}

The above properties in fact hold for {\it any} model of the singlet state satisfying no signaling and measurement independence.  That is, if only determinism (or outcome independence) is relaxed, then it must be relaxed completely, to model all singlet state correlations.  

In particular, a strong result by Branciard et al. states that any underlying model of the singlet state with $S=M=0$ must almost always predict a 50:50 chance of spin up or down in any direction, i.e, 
\[ p(a|x,\lambda)= \frac{1}{2} = p(b|y,\lambda) \]
for all $\lambda$, except possibly on a set of total probability zero \cite{branc}.  This immediately implies  via Eqs.~(\ref{idef}) and (\ref{crandbound}) that $I=1/2$ and $C_{random}=1$ bit, as claimed.  It further implies, using the notation of Eq.~(\ref{cmn}), that the joint probability distribution $p(a,b|x,y,\lambda)$ is of the form $(c_\lambda,1/2-c_\lambda,1/2-c_\lambda,c_\lambda)$ for almost all underlying variables, with $0\leq c_\lambda\leq 1/2$ (note that the singlet state correlation in Eq.~(\ref{singlet}) is also of this form).  But for the case $x=y$ one has, via Eqs.~(\ref{underlying}) and (\ref{singlet}),
\begin{eqnarray*} 
p(a=b|x,x)=0&=&\int d\lambda \,p(a=b|x,x,\lambda)\,p(\lambda|x,x) \\
&=& 2\int d\lambda\, c_\lambda \, p(\lambda|x,x) .
\end{eqnarray*}
Hence, $c_\lambda=0$ for this case with probability unity, i.e., the joint distribution is of the form $(0,1/2,1/2,0)$.  It immediately follows via Eqs.~(\ref{odef}), (\ref{oi}) and (\ref{obound}) that $O=1$ and $C_{outcome}=1$ bit, as claimed.

\subsection{Relaxing no signaling}

The class of singlet state models which only relax no signaling, with $I=M=0$, are represented by the Toner-Bacon model \cite{toner}.  As noted in Sec.~IV~C, this model in fact has the maximal possible degree of signaling, i.e,
\begin{equation} \label{stb}
 S^{TB}=1,~~~~C^{TB}_{sig}=1 {\rm ~bit}. 
 \end{equation}
These properties in fact hold for all deterministic measurement independent models of the singlet state, and hence the Toner-Bacon model is a canonical representative of such models.  

To demonstrate the generic nature of Eq.~(\ref{stb}) for $I=M=0$, note first from Eq.~(\ref{imin}) that for deterministic underlying models one must either have $S=0$ or $S=1$.  But there are no singlet state models having $I=S=M=0$ \cite{bell}.  Hence, $S=1$, as claimed.  This immediately implies that there is some particular underlying variable, $\lambda$, for which the marginal underlying probability of one observer shifts between the values of  $0$ and $1$, in dependence on which one of two measurement settings is selected between by the other observer.  Selecting between these settings with equal prior probabilities allows transmission of $1$ bit of information per measurement, in agreement with Eq.~(\ref{siginf}).  Since this is the maximum possible for two-valued measurement outcomes, if follows that $C_{sig}=1$ bit, as claimed.

\subsection{Relaxing measurement independence}

It is seen from the above that, when relaxed individually, determinism or no signaling must be completely relaxed to model the singlet state (as must outcome independence).  It has recently been conjectured that, when {\it jointly} relaxed, the degrees of indeterminism and signaling must satisfy the complementarity relations \cite{bis,kar}
\begin{equation}
S+2I\geq 1,~~~~C_{random}+C_{sig}\geq 1~{\rm bit} .
\end{equation}
Thus, it appears that at least 1 bit of total resources is required for any measurement independent model of the singlet state.  In contrast, if instead measurement independence is relaxed, only 1/15 of a bit is required, as will be shown below. Measurement dependence is, therefore, a relatively strong resource for simulating quantum correlations.  

In particular, for $I=S=0$, a singlet state model has been recently given with  deterministic local outcomes $a={\rm sgn}\,x\cdot\lambda$ and $b=-{\rm sgn}\, y\cdot\lambda$, for measurement directions $x$ and $y$, where $\lambda$ denotes a unit 3-vector with probability density \cite{free}
\begin{eqnarray} \nonumber
p(\lambda|x,y) &:=& \frac{1+x\cdot y}{8(\pi - \phi_{xy})}  {\rm ~~~for~~} {\rm sgn~} x\cdot\lambda  = {\rm sgn~} y\cdot\lambda,\\ \label{rho}
&:=& \frac{1-x\cdot y}{8 \phi_{xy}} {\rm ~~~for~~} {\rm sgn~} x\cdot\lambda  \neq {\rm sgn~} y\cdot\lambda .
\end{eqnarray}
Here $\phi_{xy}\in [0,\pi]$ denotes the angle between these directions, and the density is defined to be zero when the denominators vanish.  The degree of measurement dependence for this model is given by \cite{free}
\begin{equation}
 M_{singlet} = 2(\sqrt{2}-1)/3 \approx 0.276 ,
\end{equation}
corresponding to a fraction of measurement independence $F_{singlet}\approx 86\%$ in Eq.~(\ref{f}).  It will be shown in Sec.~VII that these are, respectively, the smallest possible and largest possible values of $M$ and $F$, for any deterministic nonsignaling model of the singlet state.  Hence this model is minimal, with a degree of relaxation of only $14\%$ of measurement independence required.

To calculate the corresponding measurement dependence capacity $C_{meas~dep}$ in Eq.~(\ref{cdep}), note first that the entropy of the probability density $p(\lambda|x,y)$ is given by
\begin{eqnarray*}
 H_{xy}(\Lambda) &=& h(\frac{1+x\cdot y}{2}) + \frac{1}{2}(1-x\cdot y)\log_2 \phi_{xy}\\
&~&~~ +\frac{1}{2}(1+x\cdot y)\log_2 (\pi-\phi_{xy}) + \log_2 4 .
 \end{eqnarray*}
This has a maximum value of $H_{max}=\log 4\pi\approx 3.65145$ bits (achieved for $x\cdot y=0,\pm 1$), and a minimum value of $H_{\min}\approx 3.58521$ bits (for $x\cdot y\approx \pm 0.9148$, corresponding to an angle $\phi_{xy}\approx$ 24 or 156 degrees).  Thus, the probability density is always very close, in the sense of entropy, to the uniform density $1/(4\pi)$, for any joint measurement setting.

It follows that the mutual information between the measurement settings and the underlying variable is given by
\begin{eqnarray*}
H(X,Y:\Lambda) &=& H(\Lambda) -\int dxdy\,p(x,y)\,H_{xy}(\Lambda)\\
&\leq& H(\Lambda) - H_{min}\\ &\leq& \log_2 4\pi - H_{min} ,
\end{eqnarray*}
where the last inequality  is an immediate consequence of the entropy of $\lambda$ being maximised by a uniform distribution on the sphere.  Moreover, the inequalities are saturated, for example, by choosing $p(x,y)$ such that $x$ is uniformly distributed on the sphere and, for each value of $x$, $y$ is uniformly distributed on the circle $x\cdot y \approx 0.9148$.  This choice immediately gives $H_{xy}(\Lambda)=H_{min}$, while the rotational symmetry of $p(\lambda|x,y)$ in Eq.~(\ref{rho}) yields $p(\lambda)=1/(4\pi)$, and hence $H(\Lambda)=\log_2 4\pi$.  The measurement dependence capacity of the model is, therefore,
\begin{equation} \label{defect}
C_{meas~dep} = \log_2 4\pi - H_{min} \approx 0.0663~{\rm bits} .
\end{equation}
This value, about $1/15$ of a bit, is seen to be relatively small in comparison to the 1 bit required when either determinism or no signaling is relaxed, as well as to the general bound of 1 bit obtained for such models by Barrett and Gisin \cite{barrett}.  

It of interest to calculate the mutual information $H(X,Y:\Lambda)$ for this model in two particular scenarios:  when the measurement settings are chosen uniformly from the unit sphere, and when the measurement settings are chosen randomly from the 4 settings corresponding to maximum violation of the Bell-CHSH inequality.  

In the first case $p(x,y)=1/(4\pi)^2$, leading via Eq.~(\ref{rho}) to $p(\lambda) = 1/(4\pi)$.  Hence,
\[ I_{uniform}(X,Y:\Lambda)=\log_2 4\pi - \langle H_{xy}(\Lambda)\rangle \approx 0.0280~{\rm bits}. \]
This value, about 1/36 of a bit, may be favourably compared to the corresponding values of $0.85$ and 0.28 bits in the corresponding models given by Barrett and Gisin \cite{barrett, 85note}. 

In the second case, the four CHSH settings $(x,y)$, $(x,y')$, $(x',y)$ and $(x',y')$ are defined by measurement directions $x,y,x',y'$ lying on a great circle, consecutively separated by 45 degrees \cite{chsh}.  One finds by straightforward calculation that $H_{xy}(\Lambda)= \log_2 \pi + H(q/3,q/3,q/3,1-q)$ for each setting, where the second term denotes the entropy of the distribution defined by its arguments and $q=(1+1/\sqrt{2})/2$.  One further finds $H(\Lambda)=\log_2 4\pi$, yielding 
\begin{eqnarray} \nonumber I_{CHSH}(X,Y:\Lambda) &=& 2 - H(q/3,q/3,q/3,1-q) \\
 \label{463}
 &\approx& 0.0463 ~{\rm bits}, 
 \end{eqnarray}
i.e., about 1/22 of a bit.  

To emphasise just how weak a degree of correlation the latter case represents, suppose that the observers make 22 independent repetitions of the CHSH experiment.  There are then $4^{22}\approx 2\times 10^{13}$ possible sequences of joint measurement settings.  Given knowledge of the corresponding sequence $\lambda_1,\lambda_2,\dots,\lambda_{22}$ of underlying variables, the number of possible measurement settings drops by just a factor of two, to $\approx 10^{13}$.  The correlation is, therefore, very subtle.  This is of obvious interest in the physical simulation of quantum cryptographic protocols via local deterministic devices.

\section{Relaxed Bell inequalities}

The previous section demonstrates that, to model the singlet state, one or more of the properties of determinism, nonsignaling and measurement independence have to be relaxed.  As noted in Appendix A, these properties must similarly be relaxed to model violations of Bell inequalities.  Since such inequalities are directly testable, the question of just how much relaxation is required, for a given degree of violation, is studied here.  The relaxation of outcome independence is also considered, in Sec.~VII~B.

\subsection{Jointly relaxing determinism, no signaling and measurement independence}

\subsubsection{Main theorem}

Let $x,x'$ and $y,y'$ denote possible measurement settings for a first and second observer, respectively, and label each measurement outcome by $\pm 1$. If $\langle XY\rangle$ denotes the average product of the measurement outcomes, for joint measurement setting $(x,y)$, then
it is well known that the Bell-CHSH inequality \cite{chsh}
\[ \langle XY\rangle + \langle XY'\rangle + \langle X'Y\rangle - \langle X'Y'\rangle   \leq 2 \]
must be satisfied if the measured correlations admit an underlying model with $I=S=M=0$.
Conversely, if this inequality is satisfied by the measured correlations, then an underlying model can be constructed such that $I=S=M=0$ \cite{fine}.

The joint degrees of relaxation, required to model any given violation of the Bell-CHSH inequality, are precisely quantified by the following `relaxed' version:

{\bf Theorem:}
If an underlying model exists, having values of indeterminism, signaling and measurement dependence of at most $I$, $S$ and $M$, respectively, then
\begin{equation} \label{bellism}
 \langle XY\rangle + \langle XY'\rangle + \langle X'Y\rangle - \langle X'Y'\rangle   \leq B(I,S,M), 
\end{equation}
with tight upper bound
\begin{eqnarray} \nonumber
B(I,S,M) &=& 4-(1-2I)(2-3M)~{\rm for}~ S< 1-2I\\ \nonumber
&~&~~~~~~~~~~~~~~~~~~~~~~~~~~~~~{\rm and~}M<2/3,\\ \label{bism}
&=& 4~~~~~~~~~~{\rm otherwise} .
\end{eqnarray}

The theorem verifies a conjecture in Ref.~\cite{bis}, where the form of $B(I,S,0)$ was obtained.  The extension to arbitrary $M$ is nontrivial, as per the proof in Appendix B.  
Noting that $B(0,0,0)=2$, the theorem reduces to the standard Bell-CHSH inequality for models satisfying determinism, no signalling and measurement independence.  

If a given value $2+V$ is measured for the lefthand side of Eq.~(\ref{bellism}), thus violating the standard Bell-CHSH inequality by an amount $V$, the theorem imposes the strong constraint
\begin{equation} \label{v}
B(I,S,M) \geq 2+V .
\end{equation}
on the joint degrees of indeterminism, signaling and measurement dependence that must be present in any corresponding model of the violation.  This constraint may be regarded as a complementarity relation for $I$, $S$ and $M$, quantifying the tradeoff required between these quantities to model a given violation.

Note that signaling is a useful resource for modeling a violation if and only if the `gap' condition
\begin{equation} \label{gap}
S\geq S_{gap} := 1-2I  
\end{equation}
is satisfied.  This corresponds to a degree of signaling sufficient for a marginal probability to shift across the gap between the subintervals $[0,I]$ and $[0,1-I]$.  This property also holds for violations of other Bell inequalities (see Sec.~VII~C).  Note further that {\it any} violation of the Bell-CHSH inequality can be modeled if $M\geq 2/3$.

\subsubsection{Example: measurement independent models}

The case $M=0$ has been extensively discussed elsewhere \cite{bis}.  For example, a measurement independent model of the maximum quantum violation, $V=2\sqrt{2}-2$ in Eq.~(\ref{v}), exists if and only if 
\begin{equation} \label{iv}
 I\geq V/4\approx  0.207 ~{\rm and/or}~S\geq 1-V/2\approx 0.586 . 
 \end{equation}
Further, the randomness and signaling capacities must satisfy
\begin{equation} C_{random} \geq  0.736~{\rm bits, and/or~} C_{sig}\geq 0.264{\rm ~bits},
\end{equation}
via Eqs.~(\ref{crandbound}) and (\ref{siginf}). Models saturating these bounds are given in the Appendix of Ref~\cite{bis}.  

It is of interest to compare these bounds with a communication model recently given by Pawlowski et al., which in the notation of this paper corresponds to the joint distributions
\[ p(a,b|x,y,\lambda)=p(a,b|x,y',\lambda)=p(a,b|x',y,\lambda)=\delta_{a\lambda}\delta_{b\lambda}, \]
\[ p(a,b|x',y',\lambda)= \left[ p(1-\delta_{a\lambda}) + (1-p)\delta_{a\lambda}\right]\delta_{b\lambda}, \]
with $\lambda=\pm 1$ and $p:=\sqrt{2}-1\approx 0.414$ \cite{paw} (for arbitrary $p\in[0,1]$, the corresponding violation of the Bell-CHSH inequality is $V=2p$). It is straightforward to calculate $I^P=S^P=p$. Hence, the model is nonoptimal in the sense that, as per Eq.~(\ref{iv}), models exist with only half the degree of indeterminism, $I=p/2\approx 0.207$, and no signaling, $S=0$ \cite{bis}.  Note, however, that the above model is outcome independent, with $O^P=0$.  

The randomness capacity follows from Eq.~(\ref{crandbound}) as $C^P_{random}=h(p)\approx 0.979$ bits.  To calculate the signaling capacity, note that for the measurement setting $x'$, a marginal probability of the first observer shifts between $0$ and $p$, independently of $\lambda$.  Hence, if the second observer chooses between settings $y$ and $y'$ with prior probabilities $w$ and $w'=1-w$, the mutual information that can be communicated is $H_\lambda(A:Y)=h(w'p)-w'\,h(p)$, with $h(x)$ as per Eq.~(\ref{hx}). For $p=\sqrt{2}-1$ this is maximised for $w'\approx 0.393$, yielding the corresponding signaling capacity 
$C^P_{sig} \approx 0.256$ bits.

To compare $C^P_{sig}$ with the communication capacity in Eq.~({\ref{ccom}), note that 
the model is implemented via the second observer sending a message bit $m=0,1$ to the first observer, with corresponding probabilities $p(m|y)=\delta_{m0}$ and $p(m|y')=(1-p)\delta_{m0}+p\delta_{m1}$, independently of the underlying variable $\lambda$ \cite{paw,cov}.  Hence, if the settings $y$ and $y'$ are chosen with prior probabilities $w$ and $w'=1-w$, the mutual information between the setting and the message is given by
\[  H_\lambda(M:Y,B)  = H(M:Y) = h(w'p)-w'\,h(p) , \]
which is equal to $H_\lambda(A:Y)$ calculated above. Hence, noting the roles of the first and second observers are reversed relative to the discussion in Sec.~IV~D, the model is nonredundant, and
\begin{equation}
 C^P_{commun} = C^P_{sig} \approx 0.256 {\rm ~bits} .
 \end{equation}
Finally, it may be noted that for the choice $w=w'=1/2$, the mutual information $H(M:Y)$ is $h(p/2)-h(p)/2\approx 0.247$ bits.  This corrects the value of $h(p/2)\approx 0.736$ bits given in Ref.~\cite{paw}.  Thus, fortuitously, less communication is required in this case than was originally thought.

\subsubsection{Example: nonsignaling models}

The class of nonsignaling models, with $S=0$, is of obvious interest.  The upper bound of the theorem in Eq.~(\ref{bellism}) reduces in this case to
\begin{eqnarray} \nonumber
B(I,0,M) &=& 4-(1-2I)(2-3M)~{\rm for}~ M<2/3\\ 
 \label{bim}
&=& 4~~~~~~~~~~~~~~{\rm otherwise} .
\end{eqnarray}

Thus, for example, a nonsignaling model exists for the maximum quantum violation, $V=2\sqrt{2}-2$, if and only if $(I,M)$ lies on or above the hyperbola
\begin{equation} \label{hyp}
 (1-2I)(2-3M) = 2-V= 4- 2\sqrt{2} 
\end{equation}
in the $IM$-plane.  This hyperbola has asymptotes $I=1/2$ and $M=2/3$, and intersects the $I$-axis at $I=V/4$ and the $M$-axis at $M=V/3$.  Hence, either $I\geq V/4\approx 0.207$ or $M\geq V/3\approx 0.276$  are sufficient (but not necessary) conditions, for a nonsignaling model of the maximum quantum violation to exist.

\subsubsection{Example: local deterministic models}

It is only recently that serious attention has been paid to the case $I=S=0$ (see Secs.~V and VI).  The corresponding underlying models are both deterministic and nonsignaling, but have some degree of correlation between the measurement settings and the underlying parameter $\lambda$.  The upper bound of the theorem reduces in this case to
\begin{equation} \label{boom}
B(0,0,M) = \min\{2+3M,4\} .
\end{equation}
This bound is saturated by the models given in Tables~I and II of Ref.~\cite{free} (see also Appendix B).  

It follows via Eq.~(\ref{v}) that a local deterministic model exists for the maximum quantum violation, $V=2\sqrt{2}-2$, if and only if $M\geq V/3\approx 0.276$.  This corresponds to a fraction $F=86\%$ of measurement independence, i.e., measurement independence need only be relaxed by 14\%. Noting that the singlet state achieves this degree of violation, it further follows that the deterministic nonsignaling model of singlet state correlations given in Ref.~\cite{free} (also discussed in Sec.~VI~C above), is optimal in that it has the smallest degree of measurement dependence possible for any such model.

\subsection{Relaxing outcome independence}

The measures $I$, $S$ and $M$ are linear with respect to the relevant probability distributions, making the explicit analytic calculation of the relaxed bound $B(I,S,M)$ a tractable problem.  It is much more difficult to obtain corresponding bounds if $I$ is replaced by the quadratic measure of outcome dependence, $O$, defined in Eq.~(\ref{odef}).  

However, for the case of models satisfying no signaling and measurement independence (i.e., $S=M=0$), one may derive the relaxed Bell-CHSH inequality
\begin{equation} \label{bellout}
 \langle XY\rangle + \langle XY'\rangle + \langle X'Y\rangle - \langle X'Y'\rangle   \leq \frac{4}{2-O} ,
\end{equation}
which holds whenever a model exists with a degree of outcome dependence no greater than $O$.

Recalling that $0\leq O\leq 1$ for two-valued outcomes, the right hand side of this inequality ranges between $2$ and $4$, and reduces to the standard Bell-CHSH inequality when outcome independence is satisfied, i.e., when $O=0$.  Moreover, it follows, for a degree of violation $V$ of the Bell-CHSH inequality, that a nonsignaling and measurement independent model exists if and only if $4/(2-O)\geq 2+V$.  In particular, for the maximum quantum degree of violation, $V=2\sqrt{2}-2$, such a model exists if and only if
\begin{equation} 
O \geq \frac{2V}{2+V} = 2-\sqrt{2} \approx 0.586 . 
\end{equation}
Further, from Eq.~(\ref{obound}) the maximum mutual information between the outcomes must be at least
\begin{equation} 
C_{outcome} \geq 1-h(\frac{2+3V}{4+2V})  \approx 0.264 {\rm ~bits} . 
\end{equation}

To obtain the relaxed Bell inequality in Eq.~(\ref{bellout}), let $\langle XY\rangle_\lambda$ denote the expectation value of the product of measurement outcomes for settings $x$ and $y$, and define
\[ E_\lambda:= \langle XY\rangle_\lambda + \langle XY'\rangle_\lambda + \langle X'Y\rangle_\lambda - \langle X'Y'\rangle_\lambda .\]
Defining the probabilities $c_j$, $m_j$ and $n_j$ as per Appendix B, one has
\[ E_\lambda = 2+2\sum_{j=1}^3(2c_j-m_j-n_j) -2(2c_4-m_4-n_4). \]
Further, the no-signaling assumption allows one to rewrite the marginals as $m:=m_1=m_2$, $m':=m_3=m_4$, $n:=n_1=n_3$, and $n':=n_2=n_4$, leading to
\[ E_\lambda = 2+4(c_1+c_2+c_3-c_4)-4(m+n). \]
Now, noting Eqs.~(\ref{ccon}) and (\ref{out2}), $c_j$ must lie between the lower and upper bounds $\max\{0,m_j+n_j-1,m_jn_j-O/4\}$ and $\min\{m_j,n_j,m_jn_j+O/4\}$.  Hence, replacing $c_j$ by its upper bound for $j=1,2,3$ and $c_4$ by its lower bound, one obtains, after some simplification, the corresponding tight inequality
 \begin{eqnarray*}
  E_\lambda &\leq& 4\left[ f(1-m,1-n,O)+f(m,n',O)+f(m',n,O)\right. \\
  &~& \left. + f(m',1-n',O)\right]-4m'-2 , 
  \end{eqnarray*}
 where $f(a,b,c):=\min\{a,b,ab+c/4\}$.
The maximum value of the right hand side over all marginal probabilities $m, m', n,n'\in [0,1]$, for a fixed degree of outcome dependence $O$, is found numerically to occur when $m'=1/2$ and $n=n'=1-O/2$.  Substituting these values into the right hand side, and maximising over $m$, yields the upper bound $4/(2-O)$, achieved for $m=3/2-1/(2-O)$.  Averaging over $\lambda$ then yields Eq.~(\ref{bellout}) as required.

For the above values of $m, m', n, n'$ one has $c_1=c_2=1-O/2$ and $c_3=c_4=1/2$, implying that a set of probability distributions saturating Eq.~(\ref{bellout}) is given by
\[ p_1=p_2 \equiv \left( 1-\frac{O}{2},\frac{1+O}{2}-\frac{1}{2-O},0,\frac{1}{2-O}-\frac{1}{2} \right) , \]
\[ p_3\equiv \left( \frac{1}{2},0,\frac{1-O}{2}, \frac{O}{2} \right),~~ p_4 \equiv \left( \frac{1-O}{2}, \frac{O}{2},\frac{1}{2},0, \right), \]
where it is recalled from Appendix B that $p_1\equiv p(a,b|x,y,\lambda)$, $p_2\equiv p(a,b|x,y',\lambda)$, etc.    This model is nonsignaling by construction, but is maximally indeterministic, with $I=1/2$. Note that the distributions correspond to a PR-box for $O=1$ \cite{pr}.

The corresponding outcome correlation capacity of this model follows via Eq.~(\ref{cresidual}) as
\begin{eqnarray*} 
C_{outcome} &=& g[O/2] + g[3/2-1/(2-O)] \\&~&~- g[(1+O)/2-1/(2-O)] ,
\end{eqnarray*}
where $g[x]:=-x\log_2 x$, and ranges from a minimum of 0 for $O=0$ to a maximum of 1 bit for $O=1$.  For the case of maximum quantum violation, $O=2-\sqrt{2}$, one has $C_{outcome}\approx 0.480$ bits.  Thus, less than half a bit of outcome correlation is required to model this degree of violation.

It is possible, in principle, to generalise Eq.~(\ref{bellout}) to obtain a relaxed Bell inequality corresponding to jointly relaxing both outcome independence and no signaling. The $m_j$ and $n_j$ now remain distinct, and subject to Eq.~(\ref{sigcon}).  The corresponding bound, $B(O,S)$, would quantify the complementary contributions required from jointly relaxing outcome independence and no signaling, to model a given violation of the standard Bell-CHSH inequality.

\subsection{Relaxing $I_{3322}$ and other Bell inequalities}

Cconsider a Bell inequality of the general linear form
\[ A_{\alpha}:=\sum_{a,b,j,k} \alpha^{ab}_{jk} p(a,b|x_j,y_k)  \leq B_\alpha, \]
where the upper bound holds for any underlying model with $I=S=M=0$.  It is not difficult, in principle, to quantify the joint degrees of relaxation of determinism and no signaling required for modeling violations of such Bell inequalities.  This is done via determining the corresponding least upper bound, $B_\alpha(I,S)$, of $A_\alpha$.  

In particular, determining $B_\alpha(I,S)$ may be reduced to a standard linear programming problem  (solvable in polynomial time).  One defines the linear function $A_\alpha(\lambda)$, by replacing $p(a,b|x_j,y_k)$ with $p(a,b|x_j,y_k,\lambda)$ in the above expression  for $A_\alpha$, and maximises over all joint probability distributions subject to the linear constraints of positivity, normalisation, $p(a|x_j,y_k,\lambda),p(b|x_j,y_k,\lambda)\in [0,I]\cup[1-I,1]$, and $|p(a|x_j,y_k,\lambda)-p(a|x_j,y_{k'},\lambda)|, |p(b|x_j,y_k,\lambda)-p(b|x_j,y_{k'},\lambda)|\leq S$.  The maximum value is the desired upper bound $B_\alpha(I,S)$.  In particular, since $p(\lambda|x_j,y_k)\equiv p(\lambda)$ for $M=0$, the integration of $A_\alpha(\lambda)$ over $\lambda$ yields the relaxed Bell inequality
\[ A_\alpha \leq B_\alpha(I,S) . \]
The case where measurement independence is also relaxed is more difficult (see, eg, Appendix~B for the case of the relaxed Bell-CHSH inequality), and a general procedure remains to be found.

As an example which can be treated analytically, a variant of the $I_{3322}$ inequality obtained by Collins and Gisin will be considered here.  The $I_{3322}$ inequality is the canonical Bell inequality for the case of 3 measurement settings for each observer and two-valued measurement outcomes, and has the form \cite{collins}
\begin{eqnarray*}
 I_{3322}(a,b) &:=& \sum_{j,k=1}^3 \alpha_{jk} p(a,b|x_j,y_k) - p(a|x_1)  \\
 &~&~ -  2 p(b|y_1) -p(b|y_2) \leq 0, 
 \end{eqnarray*}
with $\alpha_{jk}= 1$ for $j+k\leq 4$, $\alpha_{23}=\alpha_{32}=-1$, and $\alpha_{33}=0$.  
Note that this form is not suitable for dealing with models having a non-zero degree of signaling $S$, since the marginals $p(a|x_j)$ and $p(b|y_j)$ are not well defined in such a case (eg, one may have $p(a|x_j,y_1,\lambda)\neq p(a|x_j,y_2,\lambda)$).  However,  multiplying by the nonnegative quantity $1+ab$ and summing over $a$ and $b$ yields a suitable variant:
\begin{equation} \label{bella}
 A_{3322}:= \sum_{j,k} \alpha_{jk}\, \langle X_jY_k\rangle \leq 4 ,
 \end{equation}
where $\langle X_jY_k\rangle$ denotes the expectation of the product of measurement outcomes for the joint measurement setting $(x_j,y_k)$.

The corresponding relaxed Bell inequality is then
\begin{eqnarray} \nonumber 
A_{3322} \leq B_{3322}(I,S) &:=& 4+8I ,~ S<1-2I,\\ 
 \label{a3322}
&=& 8~~~~~~~~~~{\rm otherwise} ,
\end{eqnarray}
and is derived in Appendix C.  This inequality is tight; reduces to Eq.~(\ref{bella}) for $I=S=0$; and is seen to be exactly twice the upper bound, $B(I,S,M)$, of the relaxed Bell-CHSH inequality in Eq.~(\ref{bellism}) for $M=0$.  

A generalisation of Eq.~({\ref{a3322}) to $m$ measurement settings on each side is conjectured in Appendix C.

\section{How much free will do EPR-Kochen-Specker theorems need?}

The original Kochen-Specker theorem showed that one cannot consistently assign any pre-existing measurement outcomes to a particular set of (117) quantum observables on a three-dimensional Hilbert space, under the assumption of `noncontextuality', i.e., that the outcome assigned to one observable is independent of whether or not it is simultaneously measured with a compatible observable \cite{ks}.  A similar result was obtained independently by Bell \cite{bellrev}, but relying on a continuum of observables. Both results have the advantage of holding independently of the quantum state.  However, as pointed out by Bell, the noncontextuality assumption is rather strong.  For example, if the compatible observables are measured in the same local region of spacetime, then there is no compelling physical reason why simultaneous measurement contexts should not `interfere' with each other in some way \cite{bellrev}. 

Heywood and Redhead were able to substantially strengthen the basis for the noncontextuality assumption, by only requiring that it hold for observables measured in spacelike separated regions, and restricting attention to quantum states for which these observables were perfectly correlated \cite{heywood}. Thus, they were able to effectively replace (or justify) noncontextuality, in their version of the Kochen-Specker theorem, via the physically more plausible assumption of no signaling - albeit at the mild expense of having to restrict attention to particular quantum states.   Note also that, as per the argument for `elements of reality' by Einstein, Podolsky and Rosen (EPR) \cite{epr}, perfect correlations between distant observables motivate why one might wish to assign pre-existing measurement outcomes in the first place (see also Sec.~III~A). Hence, the Heywood-Redhead result, and later simplified versions, may be referred to as `EPR-Kochen-Specker theorems'.  

EPR-Kochen-Specker theorems are seen to rely on assumptions esentially equivalent to determinism (pre-existing outcomes), and no signaling (each outcome is independent of what is measured in a spacelike separated region).  They in fact also rely on a further assumption, only first made explicit by Conway and Kochen \cite{conway}:  that experimenters can freely choose to measure any of the observables in question.  Thus, an assumption implying measurement independence is also required. All such theorems have, therefore, similar significance to Bell inequalities.  

However, EPR-Kochen-Specker theorems are distinguished from Bell inequalities in the important respect that they are not statistical in character: they show that particular correlated observables cannot be logically assigned {\it any} set of fixed outcomes, irrespective of the probabilities of these outcomes.  Hence, relaxing the assumptions of determinism or no signaling would contradict the essence of these theorems.  In contrast, it is natural to consider by how much the degree of measurement independence must be relaxed, to be able to consistently assign such a set of pre-existing measurement outcomes.

It is shown below that an EPR-Kochen-Specker theorem  due to Mermin \cite{mermin} is quite robust: one must relax measurement independence by at least 1/3 to allow pre-existing measurement outcomes to be assigned.  In contrast, the Conway-Kochen `free will' theorem \cite{conway} and a theorem due to Hardy \cite{hardy} fail if measurement independence is relaxed by only 6.5\% and 4.5\%, respectively.  

\subsection{Relaxing Mermin's theorem}

Mermin gave an EPR-Kochen-Specker theorem for three mutually spacelike separated observers, who may be labelled Alice, Bob and Charlie.  The observers conduct a joint experiment where Alice measures one of two observables  $A,A'$, Bob measures one of two observables $B,B'$, and Charlie measures one of two observables $C, C'$, with each observer's outcome labelled by $\pm 1$.  The observables are assumed to exhibit the perfect correlations
\begin{equation} \label{perf}
 \langle ABC'\rangle = \langle AB'C\rangle=\langle A'BC\rangle = 1,~ \langle A'B'C'\rangle=-1 ,
 \end{equation}
where $\langle XYZ\rangle$ denotes the expectation value of the product of the outcomes of observables $X$, $Y$ and $Z$.  Such correlations can be implemented quantum mechanically, for example, when $A,A',B,B',C,C'$ correspond to the spin-1/2 observables $\sigma^A_x, \sigma^A_y, \sigma^B_x, \sigma^B_y, \sigma^C_x, \sigma^C_y$, respectively, and the observers share the tripartite state $|\psi\rangle$ defined by the $+1$ eigenvalues of the commuting operators $\sigma^A_x\sigma^B_x\sigma^C_y$, $\sigma^A_x\sigma^B_y\sigma^C_x$ and $\sigma^A_y\sigma^B_x\sigma^C_x$ \cite{mermin}.

Mermin argued that, if the existence of an underlying nonsignaling model is assumed, `one is impelled to conclude' that the measurement outcomes are predetermined \cite{mermin}.  Of course, one is not {\it com}pelled to conclude this: determinism does not logically follow from the combination of no signaling and perfect correlations, as discussed in Sec.~III~A.
However, if the model is {\it assumed} to be deterministic, then the outcomes of $A,A',B,B,C,C'$ are fixed prior to any measurements, and may be denoted by $a,a',b,b',c,c'=\pm 1$ for any given run of the experiment. The perfect correlations then appear to imply that 
\begin{equation}   \label{val}
abc'=ab'c=a'bc=1,~~ a'b'c'=-1, 
\end{equation}
which is clearly inconsistent for any assignment of values \cite{mermin} (since the product of the first three equations gives $a'b'c'=1$).  It therefore seems that there is no deterministic nonsignaling model of the correlations.

However, the derivation of Eq.~(\ref{val}) in fact requires a further assumption, not explicitly discussed by Mermin: that Alice can always choose which one of $A$ and $A'$ to measure in each run of the experiment, and similarly for Bob and Charlie.  If this assumption is not made, it is in fact possible to construct a deterministic nonsignaling model of the correlations in Eq.~(\ref{perf}), as is demonstrated in Table~I.

\begin{table}
\caption{\label{tab:table1}A class of local deterministic models for Mermin's correlations}
\begin{ruledtabular}
\begin{tabular}{c|cccccc|cccc}
$\lambda$ &$A$ & $B$ & $C$ & $A'$ & $B'$ & $C'$ & $p_{ABC'}$ & $p_{AB'C}$ & $p_{A'BC}$ & $p_{A'B'C'}$\\ 
\hline
$\lambda_1$ & $a_1$ & $b_1$ & $c_1$ & $a_1b_1$ & $a_1c_1$ & $b_1c_1$ & $1/3$ & $1/3$ & $1/3$ & 0\\
$\lambda_2$ & $a_2$ & $b_2$ & $c_2$ & $a_2b_2$ & $a_2c_2$ & $-b_2c_2$ & $1/3$ & $1/3$ & 0 & $1/3$\\
$\lambda_3$ & $a_3$ & $b_3$ & $c_3$ & $a_3b_3$ & $-a_3c_3$ & $b_3c_3$ & $1/3$ & 0 & $1/3$ & $1/3$\\
$\lambda_4$ & $a_4$ & $b_4$ & $c_4$ & $-a_4b_4$ & $a_4c_4$ & $b_4c_4$ & 0 & $1/3$ & $1/3$ & $1/3$
\end{tabular}
\end{ruledtabular}
\end{table}

The model in Table~I has an underlying variable $\lambda$ taking four possible values, $\lambda_1,\lambda_2,\lambda_3,\lambda_4$.  For each $\lambda_j$ the corresponding measurement outcomes are deterministically and locally specified, via 12 fixed numbers $a_j,b_j,c_j=\pm 1$.  The underlying probability density, $p(\lambda|A,B,C')$, corresponding to a joint measurement of $A$, $B$ and $C'$ is denoted by $p_{ABC'}$, and similarly for the other joint measurements appearing in Eq.~(\ref{perf}).   
It is easily checked that this model reproduces the perfect correlations in Eq.~(\ref{perf}) with, eg, $\langle ABC'\rangle=\sum_j p_{ABC'}(\lambda_j)A(\lambda_j)B(\lambda_j)C'(\lambda_j)=1$.  

Hence, there is indeed a deterministic nonsignaling model for these correlations, as claimed.  However, this is at the cost of relaxing measurement independence, i.e., of introducing correlations between the measurement settings and the underlying variable (see Sec.~V).   For example, from Table~I the joint measurement of $A'$, $B'$ and $C'$ cannot be performed if the underlying variable is equal to $\lambda_1$.  

The degree of measurement dependence of the model may be calculated via Eq.~(\ref{mdep}) as $M=2/3$, corresponding to a fraction $F=2/3$ of measurement independence in Eq.~(\ref{f}).  Thus, one third of measurement independence must be given up.  The corresponding measurement dependence capacity may also be calcuated, via Eq.~(\ref{cdep}), as $C_{meas~dep}=\log_2 4/3\approx 0.415$ bits (achieved by choosing between the four possible joint measurements with equal probabilities).   Thus, less than half a bit of correlation is required between the settings and the underlying variable.

It is important to note that the above model does not simulate the Mermin state $|\psi\rangle$; nor is that the aim here.  The much more modest aim is to calculate to the degree to which measurement independence must be relaxed to overcome the conclusions of Mermin's theorem, i.e., to provide a local deterministic model of the perfect correlations in Eq.~(\ref{perf}).  However, it would certainly be of interest to generalise the local deterministic model of the singlet state in Sec.~VI~C, to find a similar optimal model for Mermin's state.

\subsection{Relaxing the `free will' theorem}

Conway and Kochen have given a theorem of the same ilk as Mermin's theorem above, the main differences being (i) only two observers are required, and (ii) the need for a further assumption such as `free will' is explicitly noted \cite{conway}.   However, it will be seen that this `free will' theorem is weaker than Mermin's theorem, in the sense that measurement independence needs only to be relaxed by 6.5\% to give a local deterministic model of the correlations.

Briefly, Conway and Kochen consider two distant observers, each of whom measures a two-valued observable labeled by members of a particular set of unit 3-vectors, with possible measurement outcomes 0 or 1. The outcomes are assumed to exhibit perfect correlations when the same measurement direction is chosen by both observers, i.e.,
\[ p(a=b|x,x) = 1 . \]
It is further assumed that the measurements corresponding to any orthogonal triple of measurement directions, $x,y,z$ say, can be performed simultaneously by either observer, and always give the outcomes $1,0,1$ in some order.  Such correlations can be implemented quantum mechanically, for example, via the observers sharing a pair of spin-1 particles in a state of total spin zero, where the observable labeled by direction $j$ corresponds to the square of the spin observable in that direction \cite{heywood, conway}.

Conway and Kochen show there is a particular set of 33 measurement directions, $D_{33}$, for which there is no underlying model of the above correlations which satisfies determinism, no signaling and measurement independence.  They conclude that particles have `exactly the same kind' of free will as experimenters, where both indeterminism and measurement independence are equated with `free will', for particles and experimenters respectively.  However, a model having 0\% indeterminism and 93.5\% measurement independence is given below.

In particular, to construct a deterministic nonsignaling model of the above correlations, note first that $D_{33}$ is minimal in the sense observed by Peres \cite{peres}: for each direction $w\in D_{33}$ there exists a corresponding function $\theta_w(x)$, from $D_{33}$ to $\{0,1\}$, such that 
\[ \theta_w(x) + \theta_w(y) + \theta_w(z) =2 \]
for any mutually orthogonal triple $(x,y,z)$ satisfying $x,y,z\neq w$.  Hence, consider a model having the underlying joint probabilities
\[ p(a,b|x,y,\lambda_w) := \delta_{a,\theta_w(x)}\, \delta_{b,\theta_w(y)}, \]
where the possible values of the underlying variable are labeled by $w\in D_{33}$.  This model is clearly deterministic and nonsignaling, and satisfies $p(a=b|x,x) = 1$ as required.  Further, by construction, the outcomes for a simultaneous measurement of any mutually orthogonal triple $(x,y,z)$ must be $1,0,1$ in some order, provided that no member of the triple is equal to $w$.  Finally,  the latter provisio may be guaranteed to hold in any actual joint measurement by defining the probability distribution of the underlying variable to be
\begin{eqnarray*} p(\lambda_w|x,y) &:=& 0, ~~~~~~~~~~~~~~~w=x~{\rm or}~w=y,\\
&:=& \frac{\delta_{xy}}{32}+\frac{1-\delta_{xy}}{31},~~{\rm otherwise}.
\end{eqnarray*}
Hence, no measurement can be made in the direction corresponding to the label of the underlying variable.

The degree of measurement dependence of the above model can be calculated via Eq.~(\ref{mdep}) as $M=4/31$, achieved for the case of joint measurements $(x,y)$, $(x',y')$ having no directions in common.  This corresponds to a fraction $F=29/31\approx 93.5\%$ of measurement independence in Eq.~(\ref{f}), i.e., measurement independence only needs to be relaxed by $\approx 6.5\%$.  The measurement dependence capacity can be estimated via Eq.~(\ref{cdep}) as
\[ C_{meas~dep} \leq H_{max}(\Lambda) - H_{min}(\Lambda) = \log_2\frac{33}{31} , \]
where the upper entropy bound follows from $\lambda_w$ taking 33 possible values, and the lower bound corresponds to any joint setting with $x\neq y$.  Thus, $\approx 0.0902$ bits - less than one tenth of one bit of correlation - is required between the underlying variable and the measurement settings.  

\subsection{Relaxing Hardy's theorem}

Finally, it is of interest to also consider a result due to Hardy, which derives an EPR-Kochen-Specker theorem having a minor statisical element \cite{hardy}.  In particular, first and second observers each measure one of two observables $U_j$ and $D_j$, where $j=1,2$ refers to the observer. Labelling the corresponding measurement outcomes by $u_j,d_j=$ 0 or 1, it is assumed that they satisfy the perfect correlations
\[ u_1u_2=0,~~d_1=1\Rightarrow u_2=1,~~d_2=1\Rightarrow u_1=1, \]
and further that the joint outcome $d_1=d_2=1$ can occur with some probability $\gamma>0$.  Such correlations can be implemented quantum mechanically via the observers sharing one of a large class of two-qubit states, providing that \cite{hardy}
\[ \gamma\leq \gamma_{max}:= (5\sqrt{5}-11)/2\approx 9\% .  \]

Hardy argues that there is no deterministic nonsignaling model of such correlations, on the grounds that such a model must predict values $d_1=1=d_2$ in at least some instances, which is incompatible with any simultaneous assignation of values of $u_1$ and $u_2$ as per the required correlations \cite{hardy}.  However, this argument makes an implicit assumption that the model is measurement independent.  If this assumption is relaxed, it is quite straightforward to write down deterministic nonsignaling models of the correlations, as is done in Table 2.

\begin{table}
\caption{\label{tab:table2}A class of local deterministic models for Hardy's correlations [note $\gamma':=(1-\gamma)/2$]}
\begin{ruledtabular}
\begin{tabular}{c|cccc|cccc}
$\lambda$ &$u_1$ & $u_2$ & $d_1$ & $d_2$ & $p_{UU}$ & $p_{UD}$ & $p_{DU}$ & $p_{DD}$\\ 
\hline
$\lambda_1$ & $a$ & $1-a$ & $0$ & $0$ & $\gamma'$ & $\gamma'$ & $\gamma'$ & $\gamma'$\\
$\lambda_2$ & $b$ & $1-b$ & $1-b$ & $b$ & $\gamma'$ & $\gamma'$ & $\gamma'$ & $\gamma'$\\
$\lambda_3$ & $0$ & $1$ & $1$ & $1$ & $\frac{\gamma}{2}$& $0$ & $\frac{\gamma}{2}$ & $\frac{\gamma}{3}$\\
$\lambda_4$ & $1$ & $0$ & $1$ & $1$ & $\frac{\gamma}{2}$ & $\frac{\gamma}{2}$ & $0$ & $\frac{\gamma}{3}$\\
$\lambda_5$ & $1$ & $1$ & $1$ & $1$ & 0 & $\frac{\gamma}{2}$ & $\frac{\gamma}{2}$ & $\frac{\gamma}{3}$
\end{tabular}\end{ruledtabular}
\end{table}

The class of models in Table 2 is defined via an underlying variable $\lambda$ taking 5 possible values $\lambda_1,\lambda_2,\dots,\lambda_5$, and corresponding deterministic outcomes specified by two numbers $a,b=0$ or $1$ (thus, there are four distinct models, corresponding to the choices of $a$ and $b$).  The underlying probability distribution $p(\lambda|U,U)$ is denoted by $p_{UU}$, and similarly for the other joint settings $(U,D)$, $(D,U)$ and $(D,D)$.  The required correlations can all be checked to hold whenever they can be measured.  For example, $u_1u_2=0$ identically except for $\lambda=\lambda_5$, but the probability of $\lambda=\lambda_5$ vanishes for the corresponding setting $(U,U)$.  

The associated degree of measurement dependence is easily calculated via Eq.~(\ref{mdep}) as $M=\gamma$, with associated fraction  of measurement independence $F=1-\gamma/2$.  Hence,  measurement independence need only be relaxed by at most $\gamma_{max}/2\approx 4.5\%$ to model the correlations.  One can also estimate the degree of correlation required between the underlying variable and the measurement settings via
\[ C_{meas~dep} \leq H_{max}(\Lambda) - H_{min}(\Lambda) = \gamma \log_2\frac{3}{2} \approx 0.585 \gamma . \]
Here the maximum entropy value corresponds to choosing between the four joint settings with equal probabilities, while the minimum value corresponds to the $(D,D)$ setting.  For $\gamma=\gamma_{max}$ this gives a bound of $\approx 0.053$ bits.

\section{Conclusions}

The main aim of this paper has been to carefully define the quantitative degrees to which certain physical properties hold for underlying models of statistical correlations (Secs~III-V), and to show how these may be applied to determine optimal singlet state models (Sec.~VI); the minimal degrees of relaxation required to simulate violations of various Bell inequalities (Sec.~VII); and the relative robustness of Kochen-Specker theorems (Sec.~VIII).   The results help to both clarify and quantify the nonclassical nature of quantum correlations, including the resources required for their simulation.

A number of possible directions for future work are suggested by the results of the paper.  
First, while the information-theoretic measures defined in Secs.~III-V quantify various resources required to simulate correlations,  little is known about the interconversion of these resources.  For example, while Barrett and Gisin show how a communication model may be converted into a measurement dependent model \cite{barrett} (see also \cite{degorre}), with $C_{commun}=C_{meas~dep}$, it is not clear how to proceed in the reverse direction.  Nor has the conjecture $C_{sig}+C_{random}\geq 1$ bit \cite{bis,kar}, for measurement independent models of singlet state correlations, yet been proved.   

Second, for signaling to be a useful resource for modeling violations of standard Bell inequalites in Eqs.~(\ref{bellism}), (\ref{a3322}) and (\ref{aconj}), the `gap' condition $S\geq 1-2I$ in Eq.~(\ref{gap}) must be satisfied .  This condition corresponds to signaling of a degree sufficient to be able to `flip' a marginal probability from $p$ to $1-p$, and it would be of interest to know whether it generalises to all Bell inequalities.  

Third, it has been seen in Secs.~VI-VIII that the relaxation of measurement independence is a remarkably strong resource for modeling quantum correlations.  For example, as per Eq.~(\ref{463}), one requires a correlation between the measurement settings and the underlying variable of only $\approx 1/22$ of a bit, to obtain a local deterministic model of the CHSH scenario.  It would be of interest to exploit such a model to simulate quantum cryptographic protocols.  It would similarly be of interest to generalise the local deterministic model of the singlet state, discussed in Sec.~VI, to find corresponding optimal models for the quantum states that generate the perfect correlations in Sec.~VIII.  Presumably, the required degree of relaxation of measurement independence will increase with Hilbert space dimension, to some saturating value $M^*\leq 2$. It is not known if $M^*<2$.

Finally, it would be of interest to generalise the relaxed Bell inequality in Eq.~(\ref{bellout}), to include the relaxation of no signaling and measurement independence, similarly to the analogous inequality in Eq.~(\ref{bellism}).  This would also allow determination of whether the model of Pawlowski et al. \cite{paw}, discussed in Sec.~VI~A, has the minimal possible degree of signaling for the case $O=M=0$.  Another reason for pursuing such a generalisation, despite the technical difficulties due to the quadratic nature of $O$ in Eq.~(\ref{odef}), is that the degrees of relaxation $O$, $S$ and $M$ are completely independent of one another, whereas the quantities $I$ and $S$ are mutually constrained via Eq.~(\ref{imin}).

{\bf Acknowledgements}  I thank N. Gisin and C. Branciard for stimulating discussions.

\appendix

\section{Determinism vs outcome independence}

As noted in Sec.~III, any set of statistical correlations admits a deterministic model if and only if it admits an outcome independent model.  A brief proof is given here. This result further implies that derivations of Bell inequalities based on outcome independence (or factorisability) are no more general than derivations based on determinism.  A proof of the relation in Eq.~(\ref{oi}), linking the measures of indeterminism $I$ and outcome dependence $O$, is also given. 

{\bf Proposition:} {\it For any set of statistical correlations $\{p(a,b|x,y)\}$, there exists an underlying model ${\cal M}$ satisfying determinism if and only if there exists an underlying model ${\cal M'}$ satisfying outcome independence. Further, these models ``commute'' with the properties of no signaling and measurement independence, i.e., ${\cal M}$ satisfies either of these properties if and only if ${\cal M'}$ does.}

{\it Proof:}  Suppose first one has a model satisfying outcome independence, as per Eq.~(\ref{out}).  Choosing some fixed ordering of the possible results, $\{a_j\}$ and $\{b_k\}$, for each measurement, define a corresponding deterministic model via: (i) the underlying variable 
\[ \tilde{\lambda}\equiv (\lambda,\alpha,\beta) , \] 
where $\alpha$ and $\beta$ take values in the interval $[0,1)$; (ii) the corresponding probability density
\[ {p}(\tilde{\lambda}|x,y) = {p}(\lambda,\alpha,\beta|x,y) := p(\lambda|x,y),  \]
for $\tilde{\lambda}$ (i.e., $\alpha$ and $\beta$ are uniformly and independently distributed over the interval $[0,1)$); and (iii)  deterministic joint probabilities $p(a_j,b_k|\tilde{\lambda})$ equal to unity if and only if
\[
\alpha \!\in \! \left[\sum_{i< j} p(a_i|x,y,\lambda)\right.\!,\left. \sum_{i\leq j} p(a_i|x,y,\lambda)\right) , \]
\[
\beta \!\in \! \left[\sum_{i< k} p(b_i|x,y,\lambda)\right.\!,\left. \sum_{i\leq k} p(b_i|x,y,\lambda)\right)  \]
are satisfied (and equal to zero otherwise). It is trivial to check that, by construction, for any pair of measurements $x$ and $y$ one then has
\[ p(a_j,b_k|x,y) = \int d\tilde{\lambda} \,p(\tilde{\lambda}|x,y)\, p(a_j|x,\tilde{\lambda})\, p(b_k|y,\tilde{\lambda}) . \]
Hence, there is a deterministic model as claimed.  Further, $p(a|x,y,\tilde{\lambda})$ and $p(a|x,y,\tilde{\lambda})$ satisfy the no-signaling conditions in Eq.~(\ref{nosig}) if and only if $p(a|x,y,\lambda)$ and $p(b|x,y,\lambda)$ do,  while ${p}(\tilde{\lambda}|x,y)$ satisfies the measurement independence condition in Eq.~(\ref{free}) if and only if $p(\lambda|x,y)$ does.  Finally, the converse is trivial, since any deterministic model is automatically an outcome independent model. In particular, dropping explicit $x$, $y$, and $\lambda$ dependence, suppose that $p(a), p(b)\in\{0,1\}$.  Then $p(a,b)$ is no greater than either of $p(a)$ and $p(b)$, implying $p(a,b)=0$ if one of the marginals vanishes.  Otherwise $p(a)=p(b)=1$, and so $1\geq p(a,b)=p(a)+p(b)-p(a\vee b) \geq p(a)+p(b)-1=1$.  Thus, $p(a,b)=p(a)\,p(b)$ in all cases, i.e., outcome independence is satisfied.  $\diamond$

The above proposition is a simple generalisation of existing results in the literature for single measurements \cite{bellrev,hall}, and can be straightforwardly further generalised to continuous ranges of measurement outcomes and more than two observers. Note that the assumed ordering means that the model is (locally) contextual \cite{bellrev,hall}.  Fine has previously used a rather different (nonlocally contextual) construction to obtain a form of the proposition for the case of four measurement pairs \cite{fine}, which can be generalised to the case of a countable set of measurement pairs \cite{fine2}.  In contrast, the above proposition applies to {\it arbitrary} sets of measurement pairs, such as spin measurements in all possible directions (and does not require no-signaling or measurement independence assumptions as per Fine).

It follows that all derivations of Bell inequalities make assumptions equivalent to, or stronger than, the existence of an underlying model satisfying determinism, no signaling and measurement independence.  This is sometimes {\it prima facie} clear \cite{bell,chsh,bell1,collins}.  While some derivations are based on measurement independence and the factorisability property $p(a,b|x,y,\lambda)=p(a|x,\lambda)\,p(b|y,\lambda)$ \cite{bellfact, jarrett}, this latter property is equivalent to the combination of outcome independence and no signaling in Eqs.~(\ref{out}) and (\ref{nosig}), which by the above proposition is equivalent to the existence of a deterministic nonsignaling model.   Finally, some derivations are based on assuming the existence of underlying joint probability distributions for counterfactual measurement settings \cite{jointprob,hall}, however, Fine has shown this is also equivalent to the existence of an underlying model satisfying determinism, no signaling and measurement independence \cite{fine}.  

To demonstrate the relation between the degrees of indeterminism and outcome dependence in Eq.~(\ref{oi}), for the case of two-valued measurements, denote the possible outcomes by $\pm 1$ and order the joint measurement outcomes as $(+,+), (+,-), (-,+), (-,-)$.  The corresponding joint probability distribution for joint measurement setting $(x,y)$ can then be written in the form 
\begin{equation} \label{cmn}
p(a,b|x,y,\lambda)\equiv
(c, m-c, n-c, 1+c-m-n), 
\end{equation}
where $m$ and $n$ denote the corresponding marginal probabilities for a $+1$ outcome.  The positivity of probability implies that
\begin{equation} \label{ccon}
 \max\{0,m+n-1\} \leq c \leq \min\{m,n\} .
 \end{equation}
The degree of outcome dependence for a particular model follows from Eq.~(\ref{odef}) as
\begin{equation} \label{out2}
 O = 4 \sup |c-mn|, 
 \end{equation}
where the supremum is over all possible triples $(c,m,n)$ generated by the model.  

Now, writing $\overline{m}=1-m$ and $\overline{n}=1-n$, Eq.~(\ref{ccon}) is equivalent to
\[ -\min\{mn,\overline{m}\,\overline{n}\} \leq c-mn \leq \min\{m\overline{n},\overline{m}n\} , \]
and hence $|c-mn|$ can be no greater than the modulus of either bound.  But the modulus of the lower bound is $mn$ for $m+n\leq 1$ and $\overline{m}\,\overline{n}$ for $\overline{m}+\overline{n}\leq 1$, with a similar result for the upper bound, yielding
\[ |c-mn| \leq \max\left\{uv|u+v\leq 1,u\in\{m,\overline{m}\}, v\in\{n,\overline{n}\}\right\} . \]
For models having a degree of indeterminism $I$, one has $m,n\in [0,I]\cup[1-I,1]$ from Eq.~(\ref{idef}).  Hence, the righthand side has a maximum of $I(1-I)$, corresponding to $u=1-v=I$ (or $1-I$). This yields $O\leq 4I(1-I)$ via Eq.~(\ref{out2}), as required.

The joint distributions achieving the maximum value of outcome dependence, $O=4I(1-I)$, follow as $(I,0,0,1-I)$, $(1-I,0,0,I)$, $(0,I,I-I,0)$, and $(0,1-I,I,0)$.  Note that these distributions are either perfectly correlated, with $p(a=b)=1$, or perfectly anti-correlated, with $p(a=-b)=1$.

\section{Proof of relaxed Bell-CHSH inequality}

To obtain Eqs.~(\ref{bellism}) and (\ref{bism}) of the theorem in Sec.~VII~A, first write the joint probability distribution for joint measurement setting $(x,y)$ as per Eq.~(\ref{cmn}). 
If $\langle XY\rangle_\lambda$ denotes the average product of the measurement outcomes, for a fixed value of $\lambda$, then $\langle XY\rangle_\lambda = 1 + 4c-2(m+n)$.  It follows from Eq.~(\ref{ccon}), noting $2\,\max(x,y) = x+y+|x-y|$, that
\begin{equation} \label{xylam}
     2|m+n-1|-1\leq \langle XY\rangle_\lambda  \leq 1-2|m-n|, 
\end{equation}
where the upper and lower bounds are attainable via suitable choices of $c$.  

It is convenient to label the four measurement settings $(x,y)$, $(x,y')$, $(x',y)$ and $(x',y')$ by 1, 2, 3 and 4, and to write $p_1\equiv p(a,b|x,y,\lambda)$, $p_2\equiv p(a,b|x,y',\lambda)$, etc., and $P_1(\lambda)\equiv p(\lambda|x,y)$, $P_2(\lambda)\equiv p(\lambda|x,y')$, etc. Defining
\begin{eqnarray*} 
T(\lambda) &:=& P_1(\lambda) \langle XY\rangle_\lambda + P_2(\lambda)\langle XY'\rangle_\lambda + P_3(\lambda) \langle X'Y\rangle_\lambda\\
&~&~ - P_4(\lambda)\langle X'Y'\rangle_\lambda ,
\end{eqnarray*}
it immediately follows via Eq.~(\ref{xylam}) that
\[
T(\lambda) \leq P_1(\lambda)+  P_2(\lambda)+P_3(\lambda)+P_4(\lambda) -2J(\lambda) ,
\] 
where
\begin{equation} \label{jlam}
 J:= P_1|m_1-n_1| + P_2|m_2-n_2| + P_3|m_3-n_3| + P_4|m_4 + n_4-1|
\end{equation}
and the upper bound is attained via the  choices $c_j=\min\{m_j,n_j\}$ for $j=1,2,3$ and $c_4=\max\{0,m_4+n_4-1\}$.  Note that $P_j$, $m_j$, $n_j$ and $c_j$ are all functions of $\lambda$. 

Hence, the quantity on the left hand side of Eq.~(\ref{bellism}) satisfies
\begin{eqnarray} \nonumber
E &:=& \langle XY\rangle + \langle XY'\rangle + \langle X'Y\rangle - \langle X'Y'\rangle\\ \label{ebound}
 &=& \int d\lambda\, T(\lambda) \leq 4 -2 \int d\lambda\, J(\lambda) .  
 \end{eqnarray}
Thus, maximising this quantity corresponds to minimising the integral of the positive quantity $J(\lambda)$ in Eq.~(\ref{jlam}).  This minimum will now be determined, subject to the constraints imposed by the statement of the theorem, i.e.,  
\begin{equation} \label{icon}
m_j,n_j \in [0,I]\cup [1-I,1] ,
\end{equation}  
\begin{equation} \label{sigcon}
 |m_1-m_2|, |m_3-m_4|, |n_1-n_3|, |n_2-n_4| \leq S,
 \end{equation}
\begin{equation} \label{mcon}
 \int d\lambda\,\left| P_j(\lambda) - P_k(\lambda)\right| \leq M.  
 \end{equation}

To proceed, suppose first that $S\geq 1-2I$.  One may then take $J(\lambda)\equiv 0$ in Eq.~(\ref{jlam}), consistently with the above constraints, via the choices $m_j=n_j=m_4=1-n_4=I$ (or $1-I$), for $j=1,2,3$.  Hence, Eq.~(\ref{ebound}) yields the tight bound $E \leq 4$ for this case, for any $P_j(\lambda)$, as per the theorem.  Equality is obtained when, for example,
\begin{equation} \label{box2}
p_1\equiv p_2\equiv p_3 \equiv (I,0,0,1-I), ~p_4\equiv (0,I,1-I,0) .
\end{equation}

Conversely, suppose that $S< 1-2I$.  From the analysis of this case for $M=0$ in Ref.~\cite{bis}, at least one of the four absolute values in Eq.~(\ref{jlam}) for $J$ must be non-zero, for each $\lambda$, with a minimum value of $1-2I$, while the other three absolute values can be chosen to vanish. For example, choosing $m_j=n_j=I$ (or $1-I$), for $j=1,2,3,4$, gives $J(\lambda)=P_4(\lambda)\,(1-2I)$.  More generally, choosing the non-vanishing absolute value to correspond to the smallest multiplier $P_j$ in Eq.~(\ref{jlam}), for each value of $\lambda$, one obtains the tight bound 
\[ J(\lambda) \geq (1-2I) \min_j \{ P_j(\lambda)\} , \]
leading via Eq.~(\ref{ebound}) to the tight bound
\[ E\leq 4 - 2(1-2I) \int d\lambda\, \min_j \{ P_j(\lambda)\} . \]
Eq.~(\ref{bellism}) immediately follows, providing that the tight bound
\begin{equation} \label{min}
 \int d\lambda\,\min_j \{ P_j(\lambda)\} \geq \max\{ 0,1 - 3M/2\}
\end{equation}
can be established.  This will now be done.

First, since $2\,\min(x,y) = x+y-|x-y|$, one has in general that
\begin{eqnarray*}
\min\{w,x,y,z\} &=& \min \left\{ \min\{w,x\},\min\{y,z\}\right\} \\
&=& \frac{1}{2}\min\{w,x\} + \frac{1}{2}\min\{y,z\} \\
&~& - \frac{1}{2}\left|\min\{w,x\}-\min\{y,z\} \right|.
\end{eqnarray*}
Suppose that $w\leq x$.  Then if $y\leq z$ the `absolute value' term above is equal to $|w-y|$, while if $y>z$, the six possible orderings $wxzy, wzxy, wzyx, zwxy, zwyx, zywx$ are easily checked to yield an absolute value term no greater than $|w-y|$ in the first 3 cases and no greater than $|x-z|$ in the second 3 cases.  It follows that
\[ \left|\min\{w,x\}-\min\{y,z\} \right| \leq |w-y| + |x-z| \]
for $w\leq x$.  But swapping $w$ with $x$ and $y$ with $z$ does not change either side, implying that this inequality also holds for $x\leq w$.  Thus, in general,
\begin{eqnarray*}
\min\{w,x,y,z\} &\geq&  \frac{1}{2}\min\{w,x\} + \frac{1}{2}\min\{y,z\}\\
&~& - \frac{1}{2}|w-y| - \frac{1}{2} |x-z| \\
&=& \frac{1}{4}(w+x+y+z) - \frac{1}{4}|w-x|\\
&~&-\frac{1}{4}|y-z| -  \frac{1}{2}|w-y| - \frac{1}{2} |x-z| .
\end{eqnarray*}
Substituting $w=P_1(\lambda), x=P_2(\lambda)$, etc., integrating over $\lambda$, and using the measurement dependence constraint in Eq.~(\ref{mcon}), then yields Eq.~(\ref{min}) as desired (noting that the left hand side of this equation is necessarily nonnegative).

It still remains to show that the bound in Eq.~(\ref{min}) is tight.  First, for $M\geq 2/3$ one needs to find suitable $P_j(\lambda)$ such that $\min_j \{P_j(\lambda)\}\equiv 0$ for all $\lambda$.  This is achieved, for example, via a model with 4 underlying variables, $\lambda_1,\dots,\lambda_4$, as per Table~II of Ref.~\cite{free}.  In particular, choosing
$P_j(\lambda_k)$ to be $p$ for $j=k$, 0 for $j+k=5$,
and $(1-p)/2$ otherwise, with $0\leq p\leq1/3$, one easily finds that $M=2-4p$, which ranges over the interval $[2/3,2]$ as desired.  Finally, for $M<2/3$, consider a model with 5 underlying variables, $\lambda_1,\dots,\lambda_5$, as per Table~I of Ref.~\cite{free}, i.e., with $P_j(\lambda_k)=1-3p$ for $k=5$, 0 for $j+k=5$, and $p$ otherwise, again with $0\leq p<1/3$.  One easily finds that $M=2p$, which ranges over the interval $[0,2/3]$, with equality in Eq.~(\ref{min}) as required.

\section{Relaxed $I_{mm22}$ inequalities}

Here the relaxed Bell inequality of Eq.~(\ref{a3322}), related to $I_{3322}$, is proved, and a generalisation to the case of $m$ measurement settings for each observer is conjectured.

It is convenient to write the joint distribution $p(a,b|x_j,y_k,\lambda)$ as per Eq.~(\ref{cmn}), with $c$, $m$ and $n$ replaced by $c_{jk}$, $m_{jk}$ and $n_{jk}$.  Eqs.~(\ref{bella}) and (\ref{xylam}) immediately imply that 
\begin{equation} \label{upper}
A_{3322}(\lambda) \leq 8- 2K ,
\end{equation}
 with equality for suitable choices of $c_{jk}$, where
\[ K:= \sum_{j+k\leq 4} |m_{jk}-n_{jk}| + |m_{23}+n_{23}-1| + |m_{32}+n_{32}-1|  . \]
Hence, the minimum possible value of $K$ must be determined, subject to the constraints $m_{jk},n_{jk}\in [0,I\cup[1-I,1]$ and $|m_{jk}-m_{jk'}|, |n_{jk}-n_{j'k}|\leq S$.

Defining $F_{jk}:=|m_{jk}-n_{jk}|$ and $G_{jk}:=|m_{jk}+n_{kj}-1|$, one has
\begin{eqnarray*}
 2K &=& [F_{11}+F_{13}+F_{21}+G_{23}] + [F_{21}+F_{13}\\
 &~&+F_{22}+G_{23}] + [F_{11}+F_{12}+F_{31}+G_{32}]\\
&~& + [F_{21}+F_{22}+F_{31}+G_{32}] .
\end{eqnarray*}
Now, each of the square bracket terms corresponds to a particular case of the quantity $J$ defined in the Appendix of Ref.~\cite{bis}, which was shown there to have a minimum value of $1-2I$ for $S<1-2I$ and $0$ otherwise, under the corresponding constraints.     But for $S<1-2I$ these minimum values are simultaneously achieved by the choices $m_{jk}=n_{jk}=I$, while for $S\geq 1-2I$ they are simultaneously achieved by choosing $m_{jk}=n_{jk}=I$ when $j+k\leq 4$, and $m_{jk}=1-n_{jk}=I$ for $j+k=5$.  Eq.~(\ref{a3322}) of the text immediately follows via Eq.~(\ref{upper}) and integration over $\lambda$.

A plausible generalisation of Eq.~(\ref{a3322}) corresponds to relaxing a variant of the more general $I_{mm22}$ Bell inequality \cite{collins}.  This inequality holds for a choice of $m$ measurement settings for each observer, with two-valued measurement outcomes, and with the general form 
\begin{eqnarray*}
I_{mm22}(a,b) &:=& \sum_{j,k=1}^m \alpha^{(m)}_{jk} p(a,b|x_j,y_k) -p(a|x_1)\\
&~& - \sum_k (m-k)\,p(b|y_k) \leq 0,
\end{eqnarray*}
where $\alpha^{(m)}_{jk}=1$ for $j+k\leq m+1$, $\alpha^{(m)}_{jk}=-1$ for $j+k= m+2$, and $\alpha^{(m)}_{jk}=0$ otherwise.

As for $I_{3322}$, the marginal probabilities in the above inequality are not well defined for a non-zero degree of signaling, and hence it is convenient to consider the variant obtained via multiplication by $1+ab$ and summation over $a,b=\pm 1$, i.e.,
\[ A_{mm22} := \sum_{j,k=1}^m \alpha_{jk}\, \langle X_jY_k\rangle \leq \frac{1}{2}m(m-1) +1 . \]
  Note that this is equivalent to the standard Bell-CHSH inequality for $m=2$.  

It is conjectured that the corresponding relaxed Bell inequality is 
\begin{equation} \label{aconj}
A_{mm22}\leq B_{mm22}(I,S) ,
\end{equation}
where
\begin{eqnarray*} 
B_{mm22}(I,S) &:=& \frac{1}{2}(m-1)(m+8I) +1 ,~ S<1-2I,\\ 
&=& \frac{1}{2}(m-1)(m+4) +1,~~~{\rm otherwise} .
\end{eqnarray*}
This reduces to Eq.~(\ref{bellism}) for $m=2$ (with $M=0$), and to Eq.~(\ref{a3322}) for  $m=3$. Note that the upper bound is obtained for $S<1+2I$ via the choice $m_{jk}=n_{jk}=I$, and for $S\geq 1-2I$ via the choices $m_{jk}=n_{jk}=I$ when $j+k\leq m+1$ and $m_{jk}=1-n_{jk}=I$ when $j+k=m+2$.


\begin{thebibliography}{99}
\bibitem{bell}  J.S. Bell, Physics {\bf 1}, 195 (1964).
\bibitem{chsh} J.F. Clauser, M.A. Horne, A. Shimony and R.A. Holt, Phys. Rev. Lett. {\bf 23}, 880 (1969).
\bibitem{bell1} eg, M. Zukowski and C. Brukner, Phys. Rev. Lett. {\bf 88} 210401 (2002); E.G. Cavalcanti
et al., Phys. Rev. Lett. {\bf 99}, 210405 (2007).
\bibitem{bellfact} eg, J.A. Clauser and M.A. Horne, Phys. Rev. D {\bf 10},  526 (1974); T. Norsen, Found. Phys. {\bf 39},  273 (2009).
\bibitem{jointprob} eg, S.L. Braunstein and C.M. Caves, Phys. Rev. Lett. {\bf 61}, 662 (1988); B.W. Schumacher, Phys. Rev. A {\bf 44},  7047 (1991).
\bibitem{collins} D. Collins and N. Gisin, Phys. Rev. A {\bf 37}, 1775 (2004).

\bibitem{ks} S. Kochen and E.P. Specker, J. Math. Mech. {\bf 17}, 59 (1967).
\bibitem{bellrev} J.S. Bell, Rev. Mod. Phys. {\bf 38}, 447 (1966).
\bibitem{heywood} P. Heywood and M.L.G. Redhead, Found. Phys. {\bf 13}, 481 (1983).

\bibitem{mermin} N.D. Mermin, Phys. Rev. Lett. {\bf 65}, 3373 (1990).
\bibitem{hardy} L. Hardy, Phys. Rev. Lett. {\bf 71}, 1665 (1993).

\bibitem{conway}  J. Conway and S. Kochen, Found. Phys. {\bf 36} 1441 (2006); J.H. Conway and S. Kochen, Notices of the AMS {\bf 56} 226 (2009).

\bibitem{branc} C. Branciard et al., Nature Physics {\bf 4}, 681 (2008).
\bibitem{free} M.J.W. Hall, Phys. Rev. Lett. {\bf 105}, 250404 (2010).
\bibitem{bis} M.J.W. Hall, Phys. Rev. A {\bf 82}, 062117 (2010).



\bibitem{jarrett}  J.P. Jarrett, No\^{u}s {\bf 18}, 569 (1984).
\bibitem{prob} E. Schr\"{o}dinger, Proc. Am. Phil. Soc. {\bf 124}, 323 (1980).

\bibitem{bohr} N. Bohr, Phys. Rev. {\bf 48}, 696 (1935).
\bibitem{bayes}  C.M. Caves, C.A. Fuchs and R. Shack, Phys. Rev. A {\bf 65}, 022305 (2002).
\bibitem{epr} A. Einstein, B. Podolsky and N. Rosen, Phys. Rev. {\bf 47}, 777 (1935).

\bibitem{manyworlds} H. Everett III, Rev. Mod. Phys. {\bf 29}, 454 (1957).
\bibitem{inote} The degree of indeterminism can equivalently be defined via the minimum variational distance between an underlying marginal distribution $P_p:=\{p,1-p\}$ and the random distribution $P_{1/2}$, i.e., $I  = \frac{1}{2} - \inf_p |p-\frac{1}{2}| = \frac{1}{2} [ 1 - \inf_p D(P_p,P_{1/2})]$, 
where $p$ ranges over all underlying marginal probabilities. 
\bibitem{pr} P. Rastall, Found. Phys. {\bf 15},  963 (1985); S. Popescu and D. Rohrlich, Found. Phys. {\bf 24},  379 (1994).
\bibitem{infnote} The mutual information, $H(K:L)$, for two jointly measured random variables $K$ and $L$, quantifies the number of bits of information obtained per member of a sequence of values of $K$, about the corresponding sequence of values of $L$, and vice versa. 

\bibitem{pinsker} A.A. Fedotov, P. Harremo\"{e}s and F. Tops\/{o}e, IEEE Trans. Inf. Theory, {\bf 49}, 1491 (2003).
\bibitem{vigier} D. D\"{u}rr, S. Goldstein and N. Zanghi, J. Stat. Phys. {\bf 67}, 843 (1992)
\bibitem{westman} A. Valentini and H. Westman, Proc. Roy. Soc. A, {\bf 461}, 253 (2005); A.F. Bennett, eprint arXiv:0908.0270 [quant-ph].
\bibitem{varnote}  An alternative degree of signaling is defined via replacing the supremums over $a$ and $b$, in $S_{1\rightarrow 2}$ and $S_{2\rightarrow 1}$, by summations.  This measure is the maximum possible variational distance between two marginal distributions due to signaling.  For the case of two-valued outcomes it is just twice the value of the measure $S$ defined in Eq.~(\ref{sig}).

\bibitem{toner} B.F. Toner and D. Bacon, Phys. Rev. Lett. {\bf 91}, 187904 (2003).
\bibitem{cov} Note that relativistic versions of communication models require a covariant time ordering, to specify the `first' observer.   This can be defined, for example, via a preferred reference frame, or by the order in which the successive backward (or forward) lightcones of a preferred `clock' trajectory intersect events in spacetime.  
\bibitem{pxyexist} As an example where probability distribution $p(x,y)$ is not well-defined, consider the `doubling' sequence of joint measurement settings defined (for given $x_1, x_2,y_1,y_2$), by one $(x_1,y_1)$ setting, two $(x_2,y_2)$ settings, four $(x_1,y_1)$ settings, eight $(x_2,y_2)$ settings, etc. In this case, the relative frequency of $(x_1,y_1)$ does not converge to some $p(x_1,y_1)$, but oscillates between 1/3 and 2/3.  Hence, the alternative forms of measurement independence given following Eq.~(\ref{free}) are not always well-defined, making Eq.~(\ref{free}) the preferred form.  Similarly, some measures of the degree of measurement dependence, such as mutual information, require $p(x,y)$ to be well-defined, and so cannot be universally applied.
\bibitem{brans} C. Brans, Int. J. Theoret. Phys. {\bf 27}, 219 (1988).
\bibitem{price} H. Price, Mind {\bf 103}, 411 (1994); H. Price, Stud. Hist. Phil. Mod. Phys. {\bf 35}, 752 (2008).
\bibitem{barrett}  J. Barrett and N. Gisin, eprint arXiv:1008.3612 [quant-ph].
\bibitem{kar}  G. Kar et al., eprint arXiv:1009.6161 [quant-ph]. 
\bibitem{85note} The value of 0.85 bits in the Barrett-Gisin model may be recognised as the quantity $H(M,\Lambda|X)$ in Eq.~(\ref{hml}) for the Toner-Bacon nonsignaling model.  This is a special case of a clever construction in which Barrett and Gisin define a new underlying variable, $\Lambda'=(M,\Lambda)$, for a given communication model, immediately implying the identity $H(\Lambda'|X,Y)=H(M,\Lambda|X,Y)$ \cite{barrett}.
\bibitem{fine} A. Fine, Phys. Rev. Lett. {\bf 48} 291 (1982).
\bibitem{paw} M. Pawlowski et al., New J Phys. {\bf 12}, 083051 (2010). 



\bibitem{peres} A. Peres, J. Phys. A {\bf 24}, L175 (1991).
\bibitem{degorre} J. Degorre, S. Laplante and J. Roland, Phys. Rev. A {\bf 72}, 062314 (2005).
 
\bibitem{hall} M.J.W. Hall, Int. J. Theoret. Phys. {\bf 27}, 1285 (1988)

\bibitem{fine2} A. Fine, J. Math. Phys. {\bf 23}, 1306 (1982), (following Eq.~(11)).  Briefly, if $a^{(m)}_j$ and $b^{(n)}_k$ denote results for measurement pair $(x_m,y_n)$, define hidden variables  $\lambda'_{j_1k_1j_2k_2\dots}:=(a^{(1)}_{j_1},b^{(1)}_{k_1},a^{(2)}_{j_2},b^{(2)}_{k_2},\dots)$; an associated distribution $\rho'(\lambda'):=\int d\lambda \rho(\lambda)$ $p(a^{(1)}_{j_1}|x_1,\lambda) $ $p(b^{(1)}_{k_1}|y_1,\lambda)$ $p(a^{(2)}_{j_2}|x_2,\lambda)$ $p(b^{(2)}_{k_2}|y_2,\lambda)\dots$; and deterministic probabilities $p(a^{(m)}_j|x_m,\lambda'):=1$ ($:=0$) when $j=j_m$ ($j\neq j_m$) for the corresponding $a^{(m)}_{j_m}$ component of $\lambda'$, and similarly for $p(b^{(n)}_k|y_n,\lambda')$.  Since  $\lambda'$ and $\rho'(\lambda')$ depend on the entirety of the particular set of measurement pairs under consideration, the model is nonlocally contextual.





\end{thebibliography}
\end{document}